\documentclass[notitlepage,aps,prl,twocolumn,superscriptaddress,amsmath,showpacs,tightenlines,longbibliography,10pt]{revtex4-1}
\usepackage{amssymb}
\usepackage{amsmath}
\usepackage{dcolumn}
\usepackage{graphicx}
\usepackage{mathrsfs}
\usepackage{subfigure}
\usepackage{booktabs}
\usepackage{color}
\usepackage{docmute}
\usepackage{hyphenat}
\usepackage{CJKutf8}
\usepackage{multirow}
\usepackage{enumerate}

\newdimen\origiwspc
\newdimen\origiwstr

\newcommand{\ket}[1]{\mbox{$|#1\rangle$}}
\newcommand{\bra}[1]{\mbox{$\langle#1|$}}
\newcommand{\average}[1]{\mbox{$\langle#1\rangle$}}

\makeatletter
\newcommand*\bigcdot{\mathpalette\bigcdot@{.5}}
\newcommand*\bigcdot@[2]{\mathbin{\vcenter{\hbox{\scalebox{#2}{$\m@th#1\bullet$}}}}}
\makeatother

\usepackage{url}
\usepackage[colorlinks]{hyperref}
\hypersetup{%
	plainpages=true,
	breaklinks=true,
	hypertexnames=false,
	pageanchor=true,
	bookmarksnumbered=true,
	colorlinks=true,
	linkcolor={blue},
	citecolor={red},
	urlcolor={blue},
	anchorcolor={black}
}

\begin{document}
	
\title{Generating Large Cats with Nine Lives:
	Long-Lived Macroscopically Distinct Superposition States in Atomic Ensembles}

\author{Wei Qin}
\affiliation{Theoretical Quantum Physics Laboratory, RIKEN Cluster
	for Pioneering Research, Wako-shi, Saitama 351-0198, Japan}

\author{Adam Miranowicz}
\affiliation{Theoretical Quantum Physics Laboratory, RIKEN Cluster
	for Pioneering Research, Wako-shi, Saitama 351-0198, Japan}
\affiliation{Institute of Spintronics and Quantum Information,
	Faculty of Physics, Adam Mickiewicz University, 61-614 Pozna\'n, Poland}

\author{Hui Jing}
\affiliation{Theoretical Quantum Physics Laboratory, RIKEN Cluster
	for Pioneering Research, Wako-shi, Saitama 351-0198, Japan}
\affiliation{Key Laboratory of Low-Dimensional Quantum Structures and Quantum Control of Ministry of Education,
	Department of Physics and Synergetic Innovation Center for Quantum Effects
	and Applications, Hunan Normal University, Changsha 410081, China}

\author{Franco Nori}
\affiliation{Theoretical Quantum Physics Laboratory, RIKEN Cluster
	for Pioneering Research, Wako-shi, Saitama 351-0198, Japan}
\affiliation{Department of Physics, The University of Michigan,
	Ann Arbor, Michigan 48109, USA}

\begin{abstract}
We propose to create and stabilize long-lived macroscopic quantum superposition states in atomic ensembles. We show that using a fully quantum parametric amplifier can cause the simultaneous decay of two atoms and, in turn, create stabilized atomic Schr\"{o}dinger cat states. Remarkably, even with modest parameters these intracavity atomic cat states can have an extremely long lifetime, up to \emph{4 orders of magnitude} longer than that of intracavity photonic cat states under the same parameter conditions, reaching \emph{tens of milliseconds}. This lifetime of atomic cat states is ultimately limited to \emph{several seconds} by extremely weak spin relaxation and thermal noise. Our work opens up a new way toward the long-standing goal of generating large-size and long-lived cat states, with immediate interests both in fundamental studies and noise-immune quantum technologies. 
\end{abstract}

\date{\today}

\maketitle

\emph{Introduction.---}Schr\"{o}dinger cat states, which are macroscopically distinct
	superposition states, express the essence of quantum mechanics. Such states are appealing not only for fundamental studies of quantum mechanics~\cite{zurek2003decoherence,haroche2013nobel}, but also for wide applications ranging from quantum metrology~\cite{kira2011quantum,pezze2018quantum} to quantum computation~\cite{ralph2003quantum,gilchrist2004schrodinger,gribbin2014computing,ofek2016extending,cai2021bosonic}. So far, a large number of approaches~\cite{agarwal1997atomic,sackett2000experimental,leibfried2005creation,ourjoumtsev2006generating,ourjoumtsev2007generation,vlastakis2013deterministically,monz201114,lau2014proposal,etesse2015experimental,bradley2019a,lu2019global,figgatt2019parallel,omran2019generation,song2019generation,wei2020verifying,PhysRevLett.126.023602} have been proposed to generate cat states. However, these cat states (especially of large size) are extremely fragile in a noisy environment, and their fast decoherence makes them impractical for applications. Thus, the ability to stabilize cat states, as an essential prerequisite for their various applications, is highly desirable.
To address this problem, a two-photon loss has  been engineered~\cite{gerry1993generation, gilles1994generation,karasik2008criteria,mirrahimi2014dynamically} and recently experimentally demonstrated~\cite{everitt2014engineering,leghtas2015confining,touzard2018coherent,lescanne2020exponential,grimm2020stabilization}. Such a nonlinear loss can protect cat states against photon dephasing~\cite{cohen2017autonomous,mirrahimi2014dynamically} but, unfortunately, {\it not} against the unavoidable single-photon loss. This implies a significantly limited cat-state lifetime. Single-photon loss has been considered to be the major source of noise in fault-tolerant quantum computation based on cat states~\cite{ralph2003quantum,gilchrist2004schrodinger,gribbin2014computing,ofek2016extending,cai2021bosonic}.  Thus, the stabilization of large-size cat states for an extended time remains challenging.

Ensembles of atoms or spins have negligible spin relaxation; consequently, their major source of noise is spin dephasing, i.e., collective dephasing, local dephasing, and inhomogeneous broadening. This motivates us to engineer the \emph{simultaneous} decay of two atoms  of an ensemble (here denoted as ``two-atom decay") and then use it to stabilize atomic cat states. Such cat states could have a very long lifetime if the two-atom decay is possible. This is because such a decay may protect these atomic cat states against spin dephasing, which is a close analogy to the mechanism of using two-photon loss to suppress photon dephasing.

However, it seems to us that the two-atom decay, which
is fundamentally different from two-photon loss, is still
lacking. To implement it, here we propose to exploit fully quantum degenerate parametric amplification. More importantly, the lifetime of the resulting atomic cat states can be made up to \emph{4 orders of magnitude} longer than that of common intracavity photonic cat states~(see Table I in~\cite{supplement}), i.e., equal superpositions of two opposite-phase coherent states.
To ensure a fair comparison, these photonic cat states need to have the same size as our atomic cat states and also suffer from single-photon loss of the same rate as given for the signal mode. With a modest cavity decay time ($\sim16$~$\mu$s), our cat-state lifetime can reach $\sim20$~ms. This is comparable to 17~ms~\cite{deleglise2008reconstruction}, which is the longest lifetime of intracavity photonic cat states to date but which was achieved with an extreme cavity decay time ($\sim0.13$~sec). As the cavity decay time increases,
our cat-state lifetime can further increase but ultimately is limited to a maximum value determined by spin relaxation and thermal noise. 
For a typical spin relaxation time $\sim40$~sec~\cite{amsuss2011cavity,grezes2014multimode}, we can predict a maximum cat-state lifetime of $\sim3$~sec. 

\begin{figure*}[t]
	\centering
	\includegraphics[width=17.0cm]{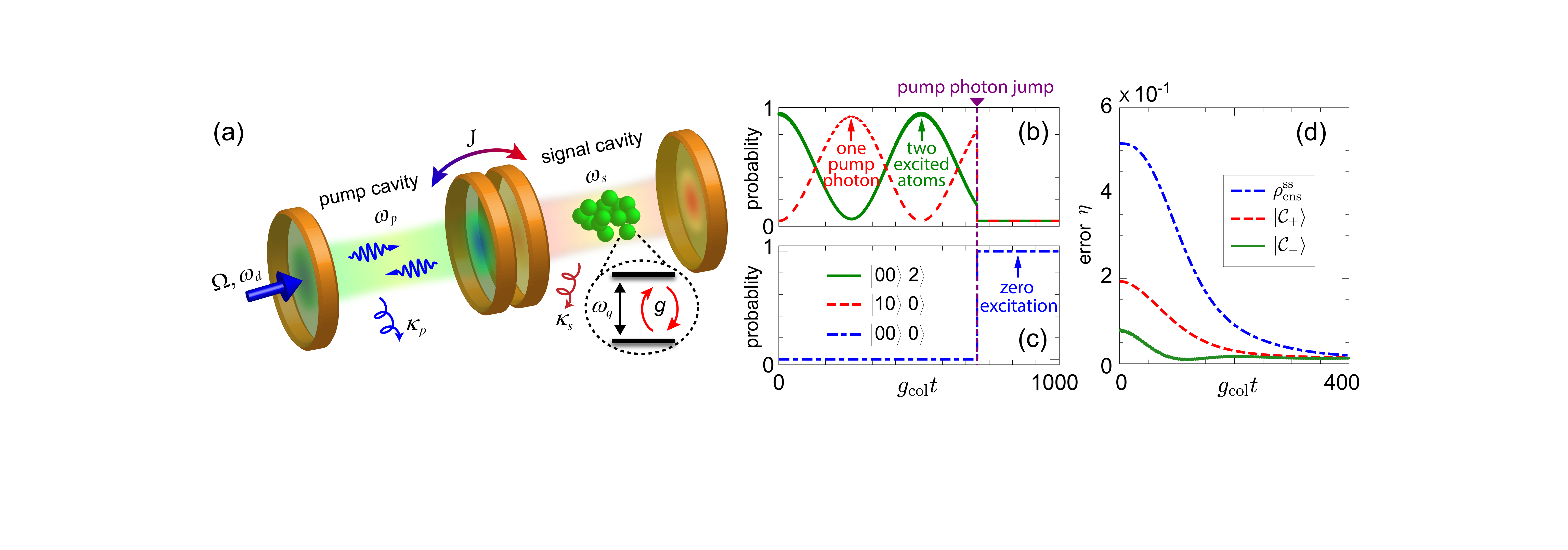}
	\caption{(a) Schematic setup of our proposal. The pump and signal cavities are coupled via a parametric coupling $J$, and the atomic ensemble is coupled to the signal cavity with a single-atom coupling $g$. The pump cavity is subject to a coherent drive with amplitude $\Omega$ and frequency $\omega_{d}$. Here, $\omega_{p}$, $\omega_{s}$ are the resonance frequencies of the pump and signal cavities, $\kappa_{p}$, $\kappa_{s}$ are their respective single-photon loss rates, and $\omega_{q}$ is the atomic transition frequency. (b), (c) Quantum Monte Carlo trajectory pictured through the probabilities of the system being in the states $\ket{m_{p}0}\ket{n}$. Initially, only two atoms in the ensemble are excited. Here,  $\kappa_{p}=0.2\chi$ and $\kappa_{s}=\Omega=0$. (d) Time evolution of the preparation error $\eta$ for a cat size $\left|\alpha\right|^{2}=1$. Here, $\kappa_{p}=5\chi$, $\kappa_{s}=0.3\kappa_p$, and the ensemble is initialized in the ground state $\ket{0}$, the single-excitation state $\ket{1}$, and a spin coherent  state $\ket{\theta_{0},0}$ with $\sqrt{N}\tan\left(\theta_{0}/2\right)=1$ for the states $\ket{\mathcal{C}_{+}}$, $\ket{\mathcal{C}_{-}}$, and $\rho_{\rm ens}^{\rm ss}$, respectively. In (b)-(d), we assume that $N=100$, $J=3g_{\rm col}$, and both cavities are initialized in the vacuum.}\label{fig_schematics}
\end{figure*}

\emph{Physical model.---}The central idea is illustrated in Fig.~\ref{fig_schematics}(a). To consider degenerate parametric amplification in the fully quantum regime, our system, inspired by recent experimental advances~\cite{leghtas2015confining,touzard2018coherent,chang2020observation,vrajitoarea2020quantum,lescanne2020exponential}, contains two parametrically coupled cavities: one as a pump cavity
with frequency $\omega_{p}$ and the other as a signal cavity with frequency $\omega_{s}$. We assume that the pump cavity is subject to a coherent drive with amplitude $\Omega$ and frequency $\omega_{d}$. The intercavity parametric coupling $J$ stimulates the conversion between pump single photons and pairs of signal photons. Furthermore, an ensemble of $N$ identical two-level atoms is placed in the signal cavity, and the atomic transition, of frequency $\omega_{q}$, is driven by a coupling $g$ to the signal photon. When $2\omega_{q}\approx\omega_{p}\ll2\omega_{s}$, a pair of excited atoms can jointly emit a pump photon. The subsequent loss of the pump photon gives rise to the two-atom decay, which in turn stabilizes large-size, extremely long-lived cat states in the ensemble.

The system Hamiltonian in a frame rotating at $\omega_{d}$ is
\begin{align}\label{eq:full Hamiltonian}
H=&\sum_{i=p,s}\delta_{i}a^{\dag}_{i}a_{i}+\delta_{q}S_{z}+J\left(a_{p}a^{\dag2}_{s}+a^{\dag}_{p}a^{2}_{s}\right)\nonumber\\
&+g\left(a_{s}S_{+}+a^{\dag}_{s}S_{-}\right)+\Omega\left(a_{p}+a^{\dag}_{p}\right),
\end{align}
where $a_{p}$, $a_{s}$ are the annihilation operators for the pump and signal modes, $S_{\pm}=S_{x}\pm iS_{y}$, $\delta_{p}=\omega_{p}-\omega_{d}$, $\delta_{s}=\omega_{s}-\omega_{d}/2$, and $\delta_{q}=\omega_{q}-\omega_{d}/2$. The collective spin operators are $S_{\alpha}=\frac{1}{2}\sum_{j=1}^{N}\sigma_{j}^{\alpha}$, with $\sigma_{j}^{\alpha}$ ($\alpha=x,y,z$) the Pauli matrices for the $j$th atom. The Lindblad dissipator, $\mathcal{L}\left(o\right)\rho=o\rho o^{\dag}-\frac{1}{2}o^{\dag}o\rho-\frac{1}{2}\rho o^{\dag}o$, describes the dissipative dynamics determined by
\begin{equation}\label{eq:full_master_equation} \dot{\rho}=-i\left[H,\rho\right]+\sum_{i=p,s}\kappa_{i}\mathcal{L}\left(a_{i}\right)\rho,
\end{equation}
where $\kappa_{p}$ and $\kappa_{s}$ are the photon loss rates of the pump and signal modes. Spin dephasing, spin relaxation, and thermal noise are discussed below.

\begin{figure*}[t]
	\centering
	\includegraphics[width=17.0cm]{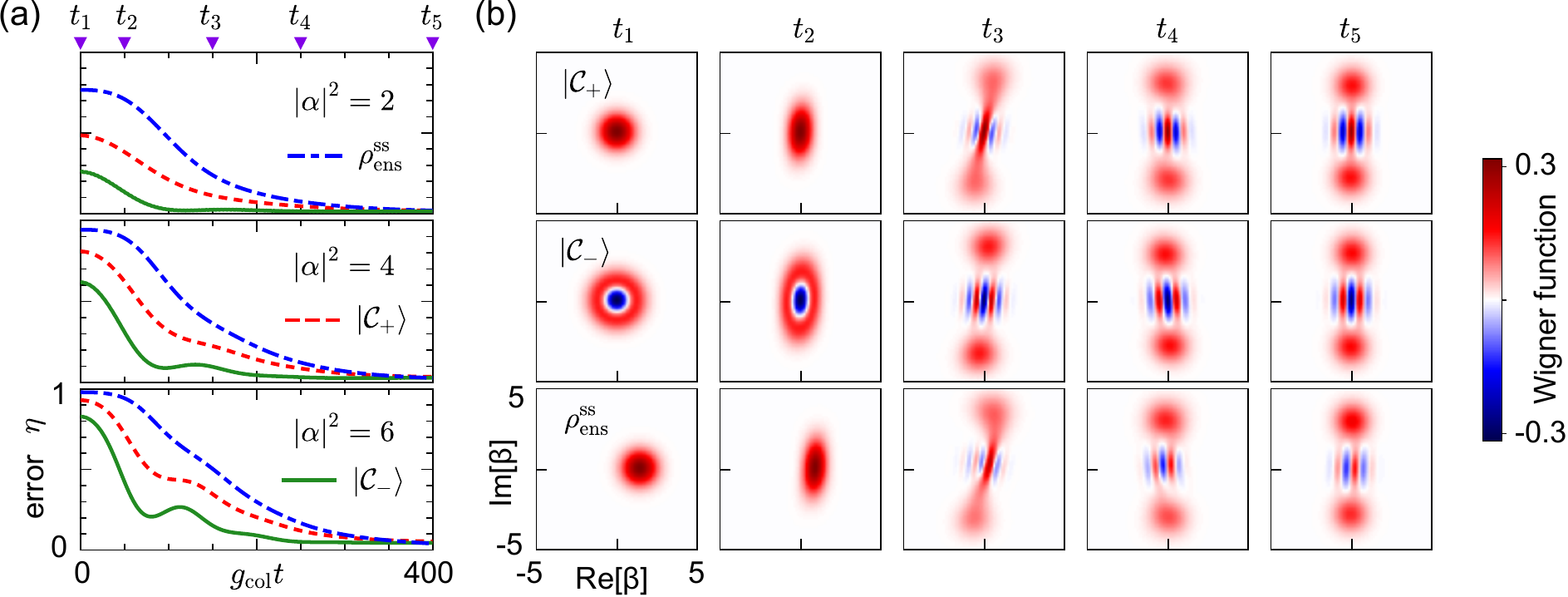}
	\caption{(a) Time evolution of the preparation error $\eta$ of the states $\ket{\mathcal{C}_{+}}$, $\ket{\mathcal{C}_{-}}$, and $\rho_{\rm ens}^{\rm ss}$ under the bosonic approximation for different cat sizes $\left|\alpha\right|^2=2$, $4$, and $6$. The initial states are chosen as in Fig.~\ref{fig_schematics}(d).  (b) Wigner function at times $t_{1},\ldots,t_{5}$ shown on top of panel (a) for the $\left|\alpha\right|^{2}=4$ cat size. The first, second, and third rows correspond to the states $\ket{\mathcal{C}_{+}}$, $\ket{\mathcal{C}_{-}}$, and $\rho_{\rm ens}^{\rm ss}$, respectively. For all plots, we set $J=3g_{\rm col}$, $\delta_{p}=J^{2}/\left(20g_{\rm col}\right)$, $\kappa_{p}=5\chi$, and $\kappa_{s}=0.3\kappa_{p}$.}\label{fig_SWA_fidel_Wigner}
\end{figure*}

We assume that $2\omega_{q}\approx\omega_{p}\approx\omega_{d}$, and the detuning $\Delta=\omega_{s}-\omega_{q}\gg\left\{g_{\rm col},J\right\}$. Here, $g_{\rm col}=\sqrt{N}g$ represents the collective coupling of the ensemble to the signal mode. Then, we can predict a parametric coupling, $\chi=g^{2}_{\rm col}J/\Delta^{2}$, between atom pairs and pump single photons. Accordingly, the Hamiltonian $H$, after time averaging~\cite{gamel2010time,shao2017generalized}, becomes
\begin{equation}\label{eq:H_prime}
H_{\rm avg}=\frac{\chi}{N}\left(a_{p}S_{+}^{2}+a^{\dag}_{p}S_{-}^{2}\right)+\Omega\left(a_{p}+a^{\dag}_{p}\right),
\end{equation}
which describes a third-order process. The stronger second-order process has been eliminated with an appropriate detuning between $\omega_{p}$ and $2\omega_{q}$ (see~\cite{supplement}). To derive $H_{\rm avg}$, we have considered the low-excitation regime, where the average number of excited atoms is much smaller than the total number of atoms. 

We now adiabatically eliminate the pump mode $a_{p}$, yielding an effective master equation
\begin{align}\label{eq:effective master equation}
\dot{\rho}_{\rm ens}=&\;-i\left[H_{\rm ens},\rho_{\rm ens}\right]\nonumber\\
&+\frac{\kappa_{\rm 1at}}{N}\mathcal{L}\left(S_{-}\right)\rho_{\rm ens}+\frac{\kappa_{\rm 2at}}{N^{2}}\mathcal{L}\left(S_{-}^{2}\right)\rho_{\rm ens},
\end{align}
where $H_{\rm ens}=i\chi_{\rm 2at}\left(S_{-}^{2}-S^{2}_{+}\right)/N$, and $\rho_{\rm ens}$ represents the reduced density matrix of the ensemble. Here, $\kappa_{\rm 2at}=4\chi^{2}/\kappa_{p}$ and $\chi_{\rm 2at}=2\Omega\chi/\kappa_{p}$ are the rates of the simultaneous decay and excitation of two atoms, respectively. Moreover, $\kappa_{\rm 1at}=\left(g_{\rm col}/\Delta\right)^2\kappa_{s}$ is the rate of the Purcell single-atom decay~(see~\cite{supplement}), and we can tune it to be $\ll\kappa_{\rm 2at}$, as long as $\kappa_{s}\ll(g_{\rm col}J/\Delta)^2/\kappa_{p}$. 

We note that the methods of Refs.~\cite{garziano2016one,macri2020spin,garziano2020atoms}
	can lead to a Hamiltonian formally similar to $H_{\rm avg}$.
	However, contrary to our method, the two-atom decay \emph{cannot}
	be realized in those methods assuming the strong cavity-photon loss.
	This is because those methods require a virtual cavity photon to mediate a third-order process, which indicates that the cavity-photon loss cannot be allowed to be strong; moreover, they also depend on a longitudinal coupling, which cannot be collectively enhanced in atomic ensembles.

To gain more insights into the engineered two-atom decay, we use the quantum Monte Carlo method~\cite{scully1997book}. In Figs.~\ref{fig_schematics}(b), (c) we plot a single quantum trajectory with the Hamiltonian $H$ and an initial state $\ket{00}\ket{2}$ (see~\cite{supplement} for more cases). Here, the first ket $\ket{m_{p}m_{s}}$ ($m_{p}, m_{s}=0,1,2,\ldots$) in the pair refers to the cavity state with $m_{p}$ pump photons and $m_{s}$ signal photons, and the second $\ket{n}$ ($n=0,1,2,\ldots$) refers to the collective spin state $\ket{S=N/2,m_{z}=-N/2+n}$, corresponding to $n$ excited atoms in the ensemble. The non-Hermitian Hamiltonian $H_{\rm NH}=H-\frac{i}{2}\kappa_{p}a_{p}^{\dag}a_{p}$ drives Rabi oscillations between $\ket{00}\ket{2}$ and $\ket{10}\ket{0}$, as shown in Fig.~\ref{fig_schematics}(b). The Rabi oscillations are then interrupted by a quantum jump $a_{p}$.  We find from Fig.~\ref{fig_schematics}(c) that the jump leaves the system in its ground state $\ket{00}\ket{0}$, implying that single-photon loss of the pump mode causes the two-atom decay.

\emph{Stabilized manifold of atomic cat states.---}When $\kappa_{\rm 1at}=0$, the dynamics of the effective master equation, Eq.~(\ref{eq:effective master equation}), describes a pairwise exchange of atomic excitations between the ensemble and its environment, thus conserving the excitation-number parity. As demonstrated in~\cite{supplement}, the ensemble is driven to an even cat state $
\ket{\mathcal{C}_{+}}=\mathcal{A}_{+}\left(\ket{\theta,\phi}+\ket{\theta,\phi+\pi}\right)$
if initialized in an even parity state, or to an odd cat state $\ket{\mathcal{C}_{-}}=\mathcal{A}_{-}\left(\ket{\theta,\phi}-\ket{\theta,\phi+\pi}\right)$
if initialized in an odd parity state. Here, $\ket{\theta,\phi}$, where $\phi=\pi/2$ and $\theta=2\arctan(\left|\alpha\right|/\sqrt{N})$, refers to a spin coherent state, and $\mathcal{A}_{\pm}=1/\{2[1\pm \exp(-2\left|\alpha\right|^{2})]\}^{1/2}$. Moreover, $\alpha=i\sqrt{\Omega/\chi}$ is the coherent amplitude. The average number of excited atoms, $\left|\alpha\right|^{2}$, of the states $\ket{\mathcal{C}_{\pm}}$ characterizes the cat size~\cite{deleglise2008reconstruction}. When assuming the initial state to be a spin coherent state $\ket{\theta_{0},\phi_{0}}$, the steady state of the ensemble is confined into a quantum manifold spanned by the states $\left\{\ket{\mathcal{C}_{+}},\ket{\mathcal{C}_{-}}\right\}$, and is expressed as
$\rho_{\rm ens}^{\rm ss}=c_{++}\ket{\mathcal{C}_{+}}\bra{\mathcal{C}_{+}}+c_{--}\ket{\mathcal{C}_{-}}\bra{\mathcal{C}_{-}}\
+c_{+-}\ket{\mathcal{C}_{+}}\bra{\mathcal{C}_{-}}+c_{+-}^{*}\ket{\mathcal{C}_{-}}\bra{\mathcal{C}_{+}}$,
where $c_{++}=\frac{1}{2}[1+\exp(-2\left|\alpha_{0}\right|^{2})]$ with $\alpha_{0}=\sqrt{N}\exp\left(i\phi_{0}\right)\tan\left(\theta_{0}/2\right)$, $c_{--}=1-c_{++}$, and $c_{+-}$ is given in~\cite{supplement}. To confirm these predictions, we numerically integrate~\cite{johansson2012qutip,johansson2013qutip2}, the master equation in Eq.~(\ref{eq:full_master_equation}), to simulate the time evolution of the preparation error $\eta=1-F$ in Fig.~\ref{fig_schematics}(d). Here, $F$ is the fidelity between the actual and ideal states. It is seen that, as expected, the ensemble states are steered into a stabilized 2D cat-state manifold with a high fidelity. 

In the low-excitation regime considered above, the collective spin in fact behaves as a quantum harmonic oscillator. This allows us to map $S_{-}$ to a bosonic operator $b$, i.e., $S_{-}\approx\sqrt{N}b$, and thus to investigate cat states of large size ($\left|\alpha\right|\geqslant2$) in large ensembles. The spin coherent state $\ket{\theta,\phi}$ accordingly becomes a bosonic coherent state $\ket{\alpha}$, such that the states $\ket{\mathcal{C}_{\pm}}$ become $
\ket{\mathcal{C}_{\pm}}=\mathcal{A}_{\pm}\left(\ket{\alpha}+\ket{-\alpha}\right)$.
With the master equation in Eq.~(\ref{eq:full_master_equation}) and under the bosonic approximation, we plot the time evolution of the preparation error $\eta$ in Fig.~\ref{fig_SWA_fidel_Wigner}(a), and the Wigner function $W\left(\beta\right)$ for different times in Fig.~\ref{fig_SWA_fidel_Wigner}(b).
We find that a cat state of size $\left|\alpha\right|^{2}=4$ is obtained after time $t\sim250/g_{\rm col}$, or more specifically, $t\sim4$~$\mu$s, for a typical collective coupling strength $g_{\rm col}/2\pi=10$~MHz~\cite{kubo2010strong, amsuss2011cavity, PhysRevA.85.012333, putz2014protecting,astner2017coherent}. 

\begin{figure}[t]
	\centering
	\includegraphics[width=8.0cm]{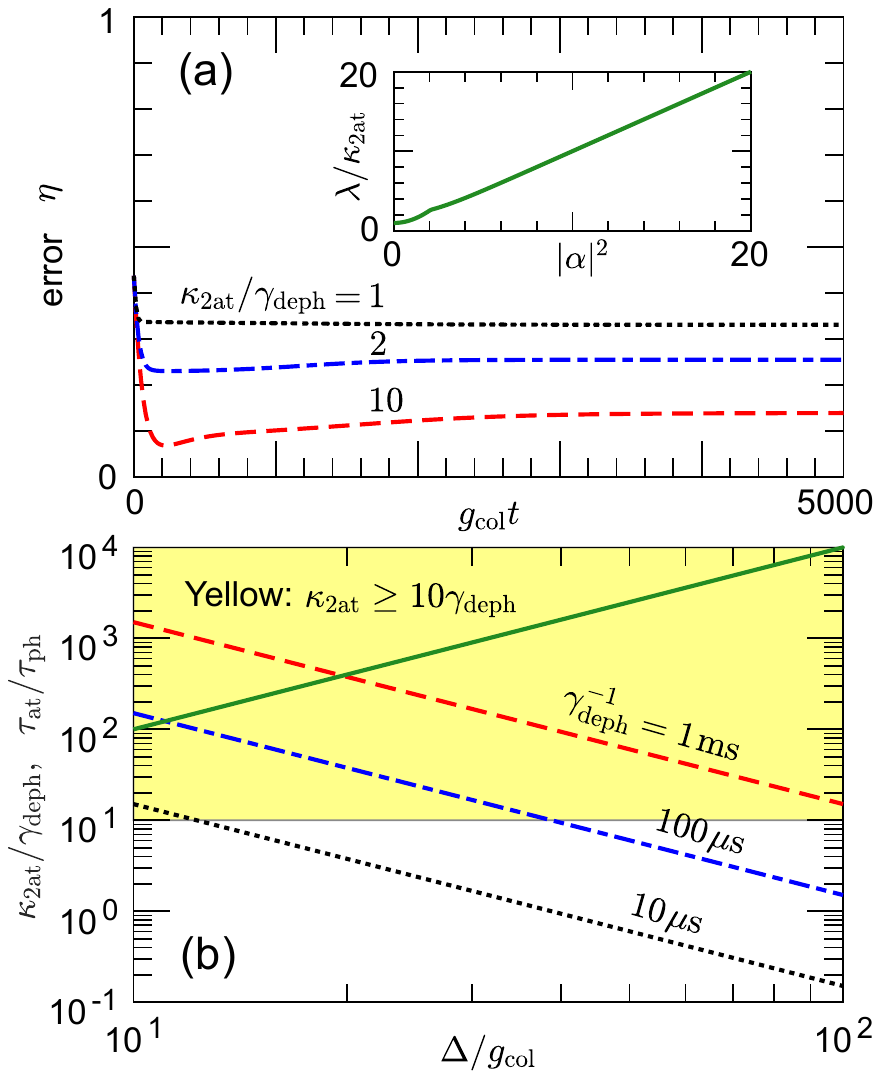}
	\caption{(a) Effects of collective dephasing, local dephasing, and inhomogeneous broadening on the preparation error $\eta$ of the state $\ket{\mathcal{C}_{+}}$ of size $|\alpha|^{2}=2$. We integrated the effective master equation, Eq.~(\ref{eq:effective master equation}), with an additional spin dephasing $\gamma_{\rm col}\mathcal{L}(S_{z})\rho_{\rm ens}$, local spin dephasing $\gamma_{\rm loc}\sum_{j=1}^{N}\mathcal{L}(\sigma_{j}^{z})\rho_{\rm ens}$, and inhomogeneous broadening $\frac{1}{2}\sum_{j=1}^{N}\delta_{j}\sigma_{j}^{z}$. The frequency shifts $\delta_{j}$ are randomly given according to a Lorentzian distribution of linewidth $\Delta_{\rm inh}$. For simplicity, we here set $N=10$, $\gamma_{\rm col}=\gamma_{\rm loc}=\Delta_{\rm inh}\equiv\gamma_{\rm deph}$ and $\kappa_{\rm 1at}=0$ so that only the effects of these dephasing processes are shown. Inset: the Liouvillian spectral gap $\lambda$ of the master equation~(\ref{eq:effective master equation}) versus the cat size $\left|\alpha\right|^{2}$ for $\kappa_{\rm 1at}=0$ under the bosonic approximation. (b) Ratio $\kappa_{\rm 2at}/\gamma_{\rm deph}$ versus the parameter $\Delta/g_{\rm col}$ for $\gamma_{\rm deph}^{-1}=10$~$\mu$s, $100$~$\mu$s, and $1$~ms for $\kappa_{p}=5\chi$ and $J/2\pi=30$~MHz. The yellow shaded area represents the $\kappa_{\rm 2at}\geq10\gamma_{\rm deph}$ regime, where spin dephasing is strongly suppressed by the two-atom decay. The solid green line shows $\tau_{\rm at}/\tau_{\rm ph}$ versus $\Delta/g_{\rm col}$. Other parameters in (a) and (b) are chosen as in Fig.~\ref{fig_SWA_fidel_Wigner}.}\label{fig_dephasing_lifetime}
\end{figure}

\emph{Suppressed spin dephasing.---} So far, we have assumed a model where there is no spin dephasing; however, there will always be some spin dephasing. Before discussing spin dephasing, let us first consider the rate $\gamma$ of convergence, i.e., how rapidly the steady cat states can be reached. To determine $\gamma$, we introduce the Liouvillian spectral gap,  $\lambda=\left|{\rm Re}\left[\lambda_{1}\right]\right|$, of the effective master equation in Eq.~(\ref{eq:effective master equation}) for $\kappa_{\rm 1at}=0$. Here, $\lambda_{1}$ is the Liouvillian eigenvalue with the smallest modulus of the real part. Since the gap $\lambda$ determines the slowest relaxation of the Liouvillian~\cite{minganti2018spectral}, we thus conclude that $\gamma>\lambda$. In the inset of Fig.~\ref{fig_dephasing_lifetime}(a), we numerically calculate the gap $\lambda$ and find $\lambda\approx\left|\alpha\right|^{2}\kappa_{\rm 2at}$ for $\left|\alpha\right|^{2}\geq2$.

Below we consider collective spin dephasing $\gamma_{\rm col}\mathcal{L}(S_{z})\rho_{\rm ens}$, local spin dephasing $\gamma_{\rm loc}\sum_{j=1}^{N}\mathcal{L}(\sigma_{j}^{z})\rho_{\rm ens}$, and inhomogeneous broadening $\frac{1}{2}\sum_{j=1}^{N}\delta_{j}\sigma_{j}^{z}$. Here, $\gamma_{\rm col}$ and $\gamma_{\rm loc}$ are the collective and local dephasing rates, respectively. Moreover, $\delta_{j}=\omega_{j}-\omega_{q}$, where $\omega_{j}$ is the transition frequency of the $j$th atom and $\omega_{q}$ can be considered as the average of transition frequencies of all the atoms~\cite{supplement}. We assume that the distribution of $\delta_{j}$ has a linewidth $\Delta_{\rm inh}$.
The three sources of dephasing noise conserve the excitation-number parity of the superradiant subspace, where the cat states are created and stabilized. Thus, all these dephasing processes can be strongly suppressed by the two-atom decay as long as $\gamma\gg \{ \gamma_{\rm col}, \gamma_{\rm loc}, \Delta_{\rm inh}\}$ (i.e., $|\alpha|^{2}\kappa_{\rm 2at}\gg\{ \gamma_{\rm col}, \gamma_{\rm loc}, \Delta_{\rm inh}\}$)~(see~\cite{supplement} for more details). Figure~\ref{fig_dephasing_lifetime}(a) shows the dependence of such a dissipative suppression on the ratio $\kappa_{\rm 2at}/\gamma_{\rm deph}$, assuming $\gamma_{\rm col}=\gamma_{\rm loc}=\Delta_{\rm inh}\equiv\gamma_{\rm deph}$. It is seen that for $\kappa_{\rm 2at}=10\gamma_{\rm deph}$, corresponding to an ensemble coherence time of $\gamma_{\rm deph}^{-1}\sim 27$~$\mu$s, a steady cat state is generated, implying a significant suppression of spin dephasing. 
We note that in Fig.~\ref{fig_dephasing_lifetime}(a) the error $\eta$
is limited by a small $N$, especially
for $\kappa_{\rm 2at}=10\gamma_{\rm deph}$, and a larger $N$ could lead to a smaller $\eta$ until the bosonic approximation is well satisfied.

\emph{Cat-state lifetime.---} Let us now consider the cat-state lifetime $\tau_{\rm at}$. According to the above discussions, the effects of spin dephasing on $\tau_{\rm at}$ can be excluded. This lifetime is thus determined by the Purcell decay rate $\Gamma_{\rm 1at}=2\left|\alpha\right|^{2}\kappa_{\rm 1at}$, such that 
\begin{equation}\label{eq:atomic cat lifetime}
\tau_{\rm at}=\Gamma_{\rm 1at}^{-1}=\left(\frac{\Delta}{g_{\rm col}}\right)^{2}\frac{1}{2\left|\alpha\right|^{2}\kappa_{s}}.
\end{equation}		
Note that intracavity photonic cat states, i.e., equal superpositions of two opposite-phase coherent states, rapidly decohere into statistical mixtures due to single-photon loss. The lifetime of such photonic cat states is thus given by $\tau_{\rm ph}=1/2\left|\alpha\right|^{2}\kappa_{s}$~\cite{haroche2006exploring}.
Here, for a fair comparison, we have assumed the same cat size $\left|\alpha\right|^{2}$ as our atomic cat states, and the same single-photon loss rate $\kappa_{s}$ as given for the signal cavity. It is seen that $\tau_{\rm at}$ is longer by a factor of $\left(\Delta/g_{\rm col}\right)^{2}$ compared to $\tau_{\rm ph}$. To make $\tau_{\rm at}/\tau_{\rm ph}$ larger, it is essential to increase $\Delta/g_{\rm col}$. However, the rate $\kappa_{\rm 2at}$, which needs to be $\gg\gamma_{\rm deph}$ as mentioned already, decreases as $\Delta/g_{\rm col}$ increases. Thus, the ratio $\Delta/g_{\rm col}$ has an upper bound for a given $\gamma_{\rm deph}$. Experimentally, the coherence time $\gamma_{\rm deph}^{-1}$ of NV-spin ensembles has reached $\sim1$~ms with spin-echo pulse sequences~\cite{hahn1950spin,stanwix2010coherence}, and if dynamical-decoupling techniques are employed, it can be even close to $1$~sec~\cite{bar2013solid}. In Fig.~\ref{fig_dephasing_lifetime}(b), the ratio $\kappa_{\rm 2at}/\gamma_{\rm deph}$ for different $\gamma_{\rm deph}$, as well as the ratio $\tau_{\rm at}/\tau_{\rm ph}$, is plotted versus $\Delta/g_{\rm col}$. Assuming a realistic parameter of $\gamma_{\rm deph}^{-1}=1$~ms, we find from Fig.~\ref{fig_dephasing_lifetime}(b) that in stark contrast to previous work on intracavity photonic cat states (see Table~I in~\cite{supplement}), our approach can lead to an increase in the cat-state lifetime of {\it up to 4 orders of magnitude} for $\kappa_{\rm 2at}\approx15\gamma_{\rm deph}$ and a very large cat size of $\left|\alpha\right|^{2}\geqslant4$. Correspondingly, for a typical single-photon loss rate of $\kappa_{s}/2\pi=10$~kHz (i.e., a cavity decay time $\sim16$~$\mu$s)~\cite{leghtas2015confining}, the lifetime of the $\left|\alpha\right|^{2}=4$ cat states resulting from our approach is $\sim20$~ms.

As the cavity loss rate $\kappa_{s}$ decreases, the lifetime $\tau_{\rm at}$ further increases and ultimately reaches its maximum value, limited by spin relaxation and thermal noise (see~\cite{supplement} for more details). This maximum lifetime is given by $\tau_{\rm at}^{\rm max}=\Gamma_{\rm relax}^{-1}$. Here, $\Gamma_{\rm relax}=[2|\alpha|^{2}(1+2n_{\rm th})+2n_{\rm th}]\gamma_{\rm relax}$~\cite{kim1992schrodinger} is the cat-state decay rate arising from spin relaxation with a rate $\gamma_{\rm relax}$ and thermal noise with a thermal average boson number $n_{\rm th}$.
For realistic parameters of $\gamma_{\rm relax}=2\pi\times4$~mHz~\cite{amsuss2011cavity,grezes2014multimode} and $T=100$~mK, we can predict a maximum lifetime of $\tau_{\rm at}^{\rm max}\sim3$~sec, which is {\it more than 2 orders of magnitude longer than the longest lifetime}, $17$~ms, of the intracavity photonic cat states reported in Ref.~\cite{deleglise2008reconstruction}.

\emph{Conclusions.---}We have introduced a method to create and stabilize large-size, long-lived Schr\"{o}dinger cat states in atomic ensembles. This method is based on the use of fully quantized degenerate parametric amplification to facilitate the simultaneous decay of two atoms, i.e., the two-atom decay. The resulting atomic cat states can last \emph{an extremely long time}, because of strongly suppressed spin dephasing and extremely weak spin relaxation and thermal noise. These long-lived cat states are promising for both fundamental tests and practical applications of quantum mechanics. Our work can further stimulate more efforts to create and protect macroscopic cat states or other fragile quantum states and to use them to improve the performance of various modern quantum technologies.

\begin{acknowledgments}
We thank Carlos S{\'a}nchez Mu{\~n}oz and Fabrizio Minganti for their valuable discussions. H.J. is supported by the
National Natural Science Foundation of China (Grants
No. 11935006 and No. 11774086) and the Science and
Technology Innovation Program of Hunan Province (Grant
No. 2020RC4047). A.M. is supported by the Polish National Science Centre (NCN) under the
	Maestro Grant No. DEC-2019/34/A/ST2/00081.
	F.N. is supported in part by:
	Nippon Telegraph and Telephone Corporation (NTT) Research,
	the Japan Science and Technology Agency (JST) [via
	the Quantum Leap Flagship Program (Q-LEAP) program,
	the Moonshot R\&D Grant No. JPMJMS2061, and
	the Centers of Research Excellence in Science and Technology (CREST) Grant No. JPMJCR1676],
	the Japan Society for the Promotion of Science (JSPS)
	[via the Grants-in-Aid for Scientific Research (KAKENHI) Grant No. JP20H00134 and the
	JSPS–RFBR Grant No. JPJSBP120194828],
	the Army Research Office (ARO) (Grant No. W911NF-18-1-0358),
	the Asian Office of Aerospace Research and Development (AOARD) (via Grant No. FA2386-20-1-4069), and
	the Foundational Questions Institute Fund (FQXi) via Grant No. FQXi-IAF19-06.
\end{acknowledgments}


%

\clearpage \widetext
\begin{center}
	\section{Supplemental Material}
\end{center}
\setcounter{equation}{0} \setcounter{figure}{0}
\setcounter{table}{0} \setcounter{page}{1} \makeatletter
\renewcommand{\theequation}{S\arabic{equation}}
\renewcommand{\thefigure}{S\arabic{figure}}
\renewcommand{\bibnumfmt}[1]{[S#1]}
\renewcommand{\citenumfont}[1]{S#1}

\section*{I\lowercase{ntroduction}}
\begin{quote}
	Here, we first compare the lifetimes of our atomic cat states and common intracavity photonic cat states. We next present how to eliminate the second-order effect and as a result to make the desired third-order effect dominant. Then, the Purcell single-atom decay induced by single-photon loss of the signal cavity is derived, and quantum Monte-Carlo trajectories of the ensemble-cavity system are shown. Furthermore, we give the detailed derivation of atomic cat states stabilized by the two-atom decay, and show the strong suppression of spin dephasing by this nonlinear two-atom decay. We then discuss the source of spin dephasing due to inhomogeneous broadening in nitrogen-vacancy center ensembles. Finally, we discuss the effects of spin relaxation and thermal noise on the cat state lifetime, and also show the maximum cat state lifetime limited by them.
\end{quote}

\subsection{S1. Comparison of the lifetimes of our atomic cat states and intracavity photonic cat states}
\label{sec:Comparison of the lifetime of intracavity photonic cat states and our atomic cat states}

\begin{table*}[b]
	\caption{Some relevant parameters of experimentally implemented intracavity photonic cat states $\ket{\mathcal{C}_{\pm}}_{\rm ph}$. Here, $\left|\alpha\right|^2$ characterizes the cat size, $T_{c}$ is the cavity photon lifetime, $\kappa_{s}=1/T_{c}$ is the cavity photon loss rate,  $\tau_{\rm exp}$ is the cat state lifetime measured in experiments, and $\tau_{\rm theor}=1/(2\left|\alpha\right|^2\kappa_{s})$ is the theoretical prediction of the cat state lifetime. For comparison, we also list at the end of the table the corresponding theoretical predictions for our atomic cat states $\ket{\mathcal{C}_{\pm}}$.}
	\vspace*{0.2cm}
	\setlength{\tabcolsep}{4.5mm}\renewcommand{\arraystretch}{1.20}{
		\begin{tabular}{|c|c|c|c|c|c|c|}
			\hline 
			Ref. & approach type &  $\left|\alpha\right|^{2}$ & $T_{c}$~($\mu$s) & $\kappa_{s}/{2\pi}$ (kHz) & $\tau_{\rm exp}$~($\mu$s) & $\tau_{\rm theor}$~($\mu$s)  \\
			\hline
			\cite{S_deleglise2008reconstruction} & unitary evolution & 3.0 & $1.3\times10^{5}$ & $1.2\times10^{-3}$ & $1.7\times10^{4}$ & $2.2\times10^{4}$  \\
			\hline	
			\cite{S_lescanne2020exponential} & reservoir engineering & 5.8 &  3.0  & 53.0 &  0.2  & 0.26   \\
			\hline
			\cite{S_vlastakis2013deterministically} & unitary evolution & 28 & 22.1 & 7.2 & --- & 0.4 \\
			\hline
			\cite{S_leghtas2015confining} & reservoir engineering & 2.4 & 20 & 8.0 & ---  & 4.1   \\
			\hline
			\cite{S_touzard2018coherent} & reservoir engineering & 5 & 92 & 1.7 & 8 & 9.2  \\
			\hline
			\cite{S_brune1996observing} & unitary evolution & 3.3 & 160 & 1.0 & 38.4 & 35   \\
			\hline
			\cite{S_wang2019quantum} & unitary evolution & 1.4 & 0.14 &  $1.1\times 10^{3}$  & --- & $5.3\times10^{-2}$  \\
			\hline
			\cite{S_assemat2019quantum} &  unitary evolution &  11.3 & $8.1\times10^{3}$ &  $2.0\times10^{-2}$ & 200 & 360  \\
			\hline		
			\cite{S_xu2020demonstration} & unitary evolution & 2 & 692 & 0.2 & --- & 173 \\
			\hline
			\multirow{2}{*}{our results} &  \multirow{2}{*}{reservoir engineering}  &  \multirow{2}{*}{4} & 16 & 10 & --- & $2\times10^{4}$ \\
			\cline{4-7}
			& & & $5.3\times10^{3}$ & $3.0\times10^{-2}$ & --- & $2\times10^{6}$ \\
			\hline			
	\end{tabular}}\label{stab:table}
\end{table*}

The cat state lifetime can be defined as the inverse cat state decoherence rate. Sec.~\hyperref[sec:Spin relaxation, thermal noise, and the maximum cat state lifetime]{S8} shows how to derive the cat state decoherence rate and then obtain the cat state lifetime. In this section, let us first compare the lifetime of intracavity atomic cat states resulting from our approach with that of common intracavity photonic cat states, under some realistic parameters. Our atomic cat states refer to superpositions of two spin coherent states, i.e.,
\begin{equation}\label{seq:atomic cat states in spin coherent states}
\ket{\mathcal{C}_{\pm}}=\;\mathcal{A}_{\pm}\left(\ket{\theta,\phi}\pm\ket{\theta,\phi+\pi}\right),
\end{equation}
Here, $\mathcal{A}_{\pm}=1/\{2[1\pm \exp(-2\left|\alpha\right|^{2})]\}^{1/2}$, and the state $\ket{\theta,\phi}$, where $\phi=\pi/2$ and $\theta=2\arctan(\left|\alpha\right|/\sqrt{N})$, is the spin coherent state that is obtained by rotating the ground state of the ensemble by an angle $\theta$ about the axis $\left(\sin\phi, -\cos\phi,0\right)$ of the collective Bloch sphere. For a large ensemble, we can apply the bosonic approximation, which maps the collective spin of the ensemble to a quantum harmonic oscillator. Under this approximation, the spin coherent states $\ket{\theta,\phi}$ and $\ket{\theta,\phi+\pi}$ become bosonic coherent states $\ket{\alpha}$ and $\ket{-\alpha}$, respectively, with coherent amplitudes $\pm\alpha$. The atomic cat states in Eq.~(\ref{seq:atomic cat states in spin coherent states}) likewise become
\begin{equation}\label{seq:atomic cat states in bosonic coherent states}
\ket{\mathcal{C}_{\pm}}=\mathcal{A}_{\pm}\left(\ket{\alpha}\pm\ket{-\alpha}\right).
\end{equation}
Furthermore, the intracavity photonic cat states refer to
\begin{equation}\label{seq:photonic cat state}
\ket{\mathcal{C}_{\pm}}_{\rm ph}=\mathcal{A}_{\pm}\left(\ket{\alpha}_{\rm ph}\pm\ket{-\alpha}_{\rm ph}\right),
\end{equation}
where $\ket{\pm\alpha}_{\rm ph}$ are the photonic coherent states with coherent amplitudes $\pm\alpha$. It is seen, from Eqs.~(\ref{seq:atomic cat states in bosonic coherent states}) and~(\ref{seq:photonic cat state}), that $\left|\alpha\right|^{2}$ is the average number of excited atoms or photons and, thus, can characterize the cat size.

In Table~\ref{stab:table}, we list some parameters of intracavity photonic cat states $\ket{\mathcal{C}_{\pm}}_{\rm ph}$ implemented in experiments.
For comparison, we also show the corresponding results of our atomic cat states $\ket{\mathcal{C}_{\pm}}$ at the end of the table. With modest parameters the lifetime of our atomic cat states is predicted to be longer, by up to \emph{four orders of magnitude}, compared to those photonic cat states under the same parameter conditions. For a modest single-photon loss rate of $\kappa_{s}/2\pi=10$~kHz (i.e., a cavity decay time of $T_{c}\sim16$~$\mu$s), the lifetime of our atomic cat states can reach $\sim20$~ms for a cat size of $\left|\alpha\right|^2=4$. This lifetime is comparable in length to that  ($\sim17$~ms) reported in Ref.~\cite{S_deleglise2008reconstruction} in Table~\ref{stab:table}, which, to our best knowledge, is the longest lifetime of intracavity photonic cat states to date. We stress that in such a comparison our cat state lifetime is achieved with a modest cavity decay time of $T_{c}\sim16$~$\mu$s. This is in stark contrast to the cat state lifetime reported in Ref.~\cite{S_deleglise2008reconstruction}, which was achieved with an extreme cavity decay time of $T_{c}=0.13$~sec. This means that our approach can stabilize (\emph{for an extremely long time}) large-size cat states, even with common setups.

When decreasing the single-photon loss rate $\kappa_{s}$, i.e., increasing the cavity decay time $T_{c}$, our atomic cat state lifetime can further increase. For example, a single-photon loss rate $\kappa_{s}/2\pi=3.0\times10^{-2}$~kHz, corresponding to a cavity decay time $T_{c}\sim5.3$~ms, results in a cat state lifetime of $\sim 2$~sec, \emph{more than two orders of magnitude longer than the lifetime}, i.e., $17$~ms, reported in Ref.~\cite{S_deleglise2008reconstruction} in Table~\ref{stab:table}. Ultimately, the maximum value of our cat state lifetime is determined by extremely weak spin relaxation and thermal noise, reaching $\sim3$~sec.

\begin{figure}[th]
	\centering
	\includegraphics[width=9.0cm]{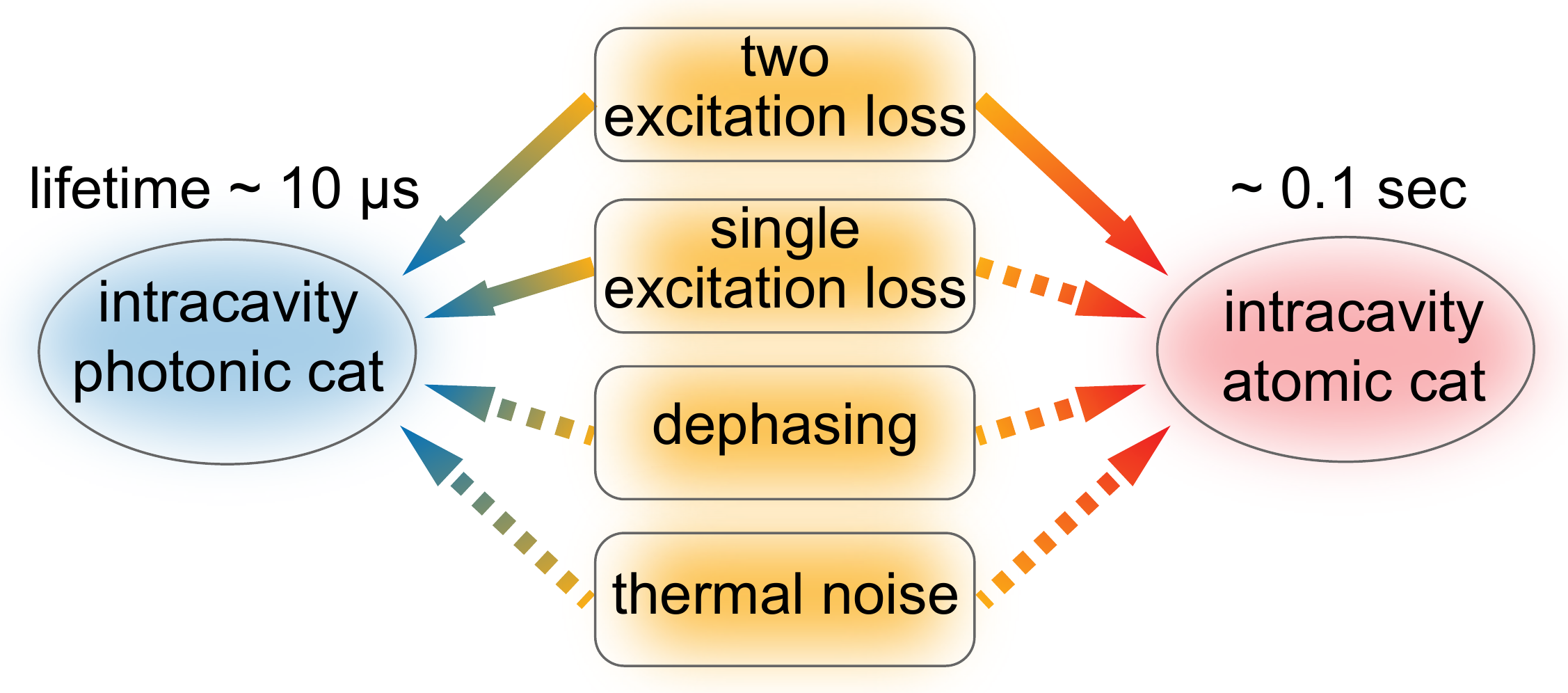}
	\caption{Comparison of the effects of noise on intracavity photonic cat stats and our atomic cat states. Solid arrows represent the strong effects, and dashed arrows represent the extremely weak or strongly suppressible effects. While the lifetime of photonic cat states is $\sim10$~$\mu$s, our atomic cat states can have a $\sim0.1$~sec lifetime.}\label{sfig_comparison}
\end{figure}

The essential reason for such an improvement in the cat state lifetime is because, as shown in Fig.~\ref{sfig_comparison}, single excitation loss of ensembles (i.e., spin relaxation) is extremely weak compared to that of cavities (i.e., single-photon loss). At the same time, spin dephasing, though stronger than photon dephasing, is greatly suppressed by the engineered two-atom decay. This is in close analogy to the mechanism of using two-photon loss to suppress photon dephasing.

\subsection{S2. Elimination of the second-order effect}
\label{sec:Compensation of second-order effects}
The time-averaged Hamiltonian $H_{\rm avg}$ in Eq.~(3) in the main article describes a third-order process, and there exists a stronger second-order process, which is described by the Hamiltonian
\begin{align}\label{eq:second order term}
H^{\left(2\right)}=-\frac{g^{2}}{\Delta}\left(2a_{s}^{\dag}a_{s}S_{z}+S_{+}S_{-}\right)
-\frac{J^{2}}{\Delta^{\prime}}\left(2a_{p}^{\dag}a_{p}-a_{s}^{\dag}a_{s}^{\dag}a_{s}a_{s}+4a^{\dag}_{p}a_{p}a_{s}^{\dag}a_{s}\right),
\end{align}
where $\Delta^{\prime}=2\omega_{s}-\omega_{p}$. In order to make the third-order  $H_{\rm avg}$ dominant, we need to eliminate the second-order $H^{\left(2\right)}$. Since the signal cavity is initialized in the vacuum state, the Hamiltonian $H^{\left(2\right)}$ is thus reduced to
\begin{equation}
H^{\left(2\right)}=-\frac{g^{2}}{\Delta}S_{+}S_{-}-\frac{2J^{2}}{\Delta^{\prime}}a_{p}^{\dag}a_{p}.
\end{equation}
We further focus our attention on the low-excitation regime, where the average number of excited atoms is much smaller than the total number of atoms. In this regime, the operator $S_{z}$ can be expressed as $S_{z}=-N/2+\delta\!S_{z}$, where $\delta\!S_{z}$ is a small fluctuation. As a result, we find
\begin{equation}
S_{+}S_{-}\approx N\delta\!S_{z},
\end{equation}
according to the identity $N\left(N/2+1\right)/2=S_{z}^{2}-S_{z}+S_{+}S_{-}$, and then obtain
\begin{equation}
H^{\left(2\right)}=-\frac{g^{2}_{\rm col}}{\Delta}\delta\!S_{z}-\frac{2J^{2}}{\Delta^{\prime}}a_{p}^{\dag}a_{p}.
\end{equation}
It is seen that the second-order process causes a Lamb shift (i.e., the first term), and a dispersive resonance shift for the pump cavity (i.e., the second term). These additional shifts can be compensated
by properly detuning the pump cavity resonance $\omega_{p}$ from twice the atomic resonance $\omega_{q}$. Hence, the second-order process can be strongly suppressed, such that the third-order process becomes dominant.

\subsection{S3. Purcell single-atom decay induced by single-photon loss of the signal cavity}
\label{sec:Single-atom decay induced by the single-photon loss of the signal cavity}
Since the signal cavity is largely detuned from both the ensemble and the pump cavity, the average number of photons inside the signal cavity is thus very low. In this case, we can only consider the vacuum state $\ket{0}$ and the single-photon state $\ket{1}$ of the signal cavity. We work within the limit where $\delta_{s}\approx\Delta\gg\left\{\delta_{p}, \delta_{q}, g_{\rm col}, J\right\}$, and the Hamiltonian in Eq.~(1) in the main article can thus be rewritten as $H=H_{e}+H_{g}+V+V^{\dagger}$. Here,
\begin{align}
H_{e}=&\;\delta_{s}\ket{1}\bra{1},\\
H_{g}=&\;\delta_{p}a^{\dag}_{p}a_{p}+\delta_{q}S_{z}+\Omega\left(a_{p}+a^{\dag}_{p}\right),
\end{align}
represents the interactions inside the excited- and ground-state subspaces, and
\begin{equation}
V=gS_{-}\ket{1}\bra{0}
\end{equation}
describes the perturbative interaction between the excited- and ground-state subspaces.
Then, according to the formalism of Ref.~\cite{S_reiter2012effective}, we can define a non-Hermitian Hamiltonian $H_{\rm NH}^{e}=H_{e}-i\kappa_{s}\ket{1}\bra{1}/2$, and obtain an effective Lindblad dissipator for the ensemble
\begin{equation}
\kappa_{s}\mathcal{L}\left[\ket{0}_{s}\bra{1}\left(H_{\rm NH}^{e}\right)^{-1}V\right]\rho_{\rm ens}=\frac{\kappa_{\rm 1at}}{N}\mathcal{L}\left(S_{-}\right)\rho_{\rm ens},
\end{equation}
where
\begin{equation}\label{seq:induced single atom decay}
\kappa_{\rm 1at}=\frac{\kappa_{s}g^{2}_{\rm col}}{\delta_{s}^{2}+\kappa_{s}^{2}/4}\approx\left(\frac{g_{\rm col}}{\Delta}\right)^{2}\kappa_{s}.
\end{equation}
This means that the single-photon loss process of the signal cavity gives rise to the single-atom decay of the ensemble. Importantly, the resulting decay rate $\kappa_{\rm 1at}$ is smaller than the cavity decay rate $\kappa_{s}$ by a factor of $\left(g_{\rm col}/\Delta\right)^{2}$. Thus, our atomic cat states have an extremely long lifetime.

\subsection{S4. Quantum Monte-Carlo trajectory for the initial states $\ket{00}\ket{3}$ and $\ket{00}\ket{4}$}
\label{sec:Quantum Monte-Carlo trajectory}

\begin{figure}[h]
	\centering
	\includegraphics[width=17.65cm]{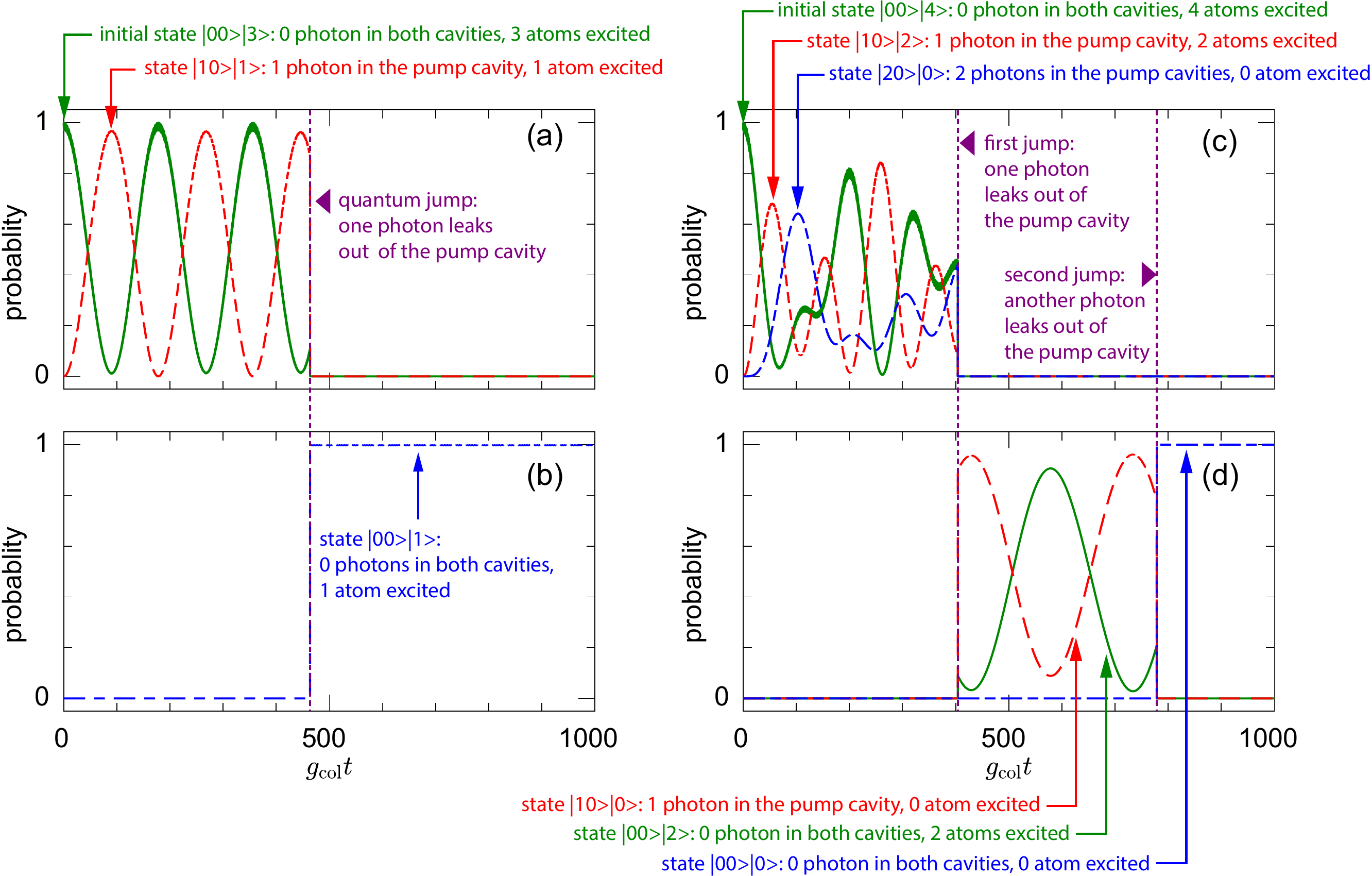}
	\caption{Quantum Monte-Carlo trajectory pictured through the probabilities of the system being in the states $\ket{m_{p}0}\ket{n}$ for the initial states (a, b) $\ket{00}\ket{3}$  and (c, d) $\ket{00}\ket{4}$. A single quantum jump $a_{p}$ gives rise to the two-atom decay in the ensemble. In all plots, we used the full Hamiltonian $H$ in Eq.~(1) in the main article, and set $N=100$, $J=3g_{\rm col}$, $\delta_{p}=J^{2}/20g_{\rm col}$, and $\kappa_{p}=0.2\chi$. In order to show more clearly the quantum jump responsible for the two-atom decay, we further set $\kappa_{s}=\Omega=0$.}\label{sfig_jump}
\end{figure}
The dynamics described by the time-averaged $H_{\rm avg}$ in Eq.~(3) of the main article implies that pairs of atoms can jointly convert their excitations into pump single photons, and then the subsequent single-photon loss process of the pump cavity results in the simultaneous decay of two atoms, i.e., the two-atom decay. 

In Fig.~\ref{sfig_jump}, we plot single quantum trajectories, utilizing the quantum Monte Carlo method, for the initial states $\ket{00}\ket{3}$ and $\ket{00}\ket{4}$. Here, the first ket $\ket{m_{p}m_{s}}$ ($m_{p}, m_{s}=0,1,2,\ldots$) in the pair refers to the cavity state with $m_{p}$ pump photons and $m_{s}$ signal photons, and the second $\ket{n}$ ($n=0,1,2,\ldots$) refers to the collective spin state $\ket{S=N/2,m_{z}=-N/2+n}$, corresponding to $n$ excited atoms in the ensemble. 

For the former case, where initially the ensemble has three excited atoms, we find from Figs.~\ref{sfig_jump}(a, b) that two excited atoms, as a pair, decay via a single-photon loss process of the DPA pump (corresponding to a quantum jump), and one excited atom is kept in the ensemble because alone it cannot emit a single photon. If there are initially four excited atoms as shown in Figs.~\ref{sfig_jump}(c, d), all excited atoms, as two pairs, can decay sequentially via two single-photon loss processes of the DPA pump (corresponding to two quantum jumps). 

\subsection{S5. Stabilized atomic cat states by the two-atom decay}
\label{sec:Stabilized atomic cat states by the two-atom decay}
In this section we show a detailed derivation of atomic cat states stabilized by the engineered two-atom decay. We begin with the effective master equation given in Eq.~(4) of the main text 
\begin{equation}
\dot{\rho}_{\rm ens}=\;i\left[\rho_{\rm ens},H_{\rm ens}\right]
+\frac{\kappa_{\rm 1at}}{N}\mathcal{L}\left(S_{-}\right)\rho_{\rm ens}+\frac{\kappa_{\rm 2at}}{N^{2}}\mathcal{L}\left(S_{-}^{2}\right)\rho_{\rm ens},
\end{equation}
Here, 
\begin{align}
H_{\rm ens}=\;&\frac{i}{N}\chi_{\rm 2at}\left(S_{-}^{2}-S^{2}_{+}\right),\\
\chi_{\rm 2at}=\;&\frac{2\Omega\chi}{\kappa_{p}},\\
\kappa_{\rm 1at}=\;&\left(\frac{g_{\rm col}}{\Delta}\right)^2\kappa_{s},\\
\kappa_{\rm 2at}=\;&\frac{4\chi^{2}}{\kappa_{p}}.
\end{align}
To proceed, we assume that $\kappa_{\rm 1at}=0$, such that the single-atom decay induced by the signal cavity is subtracted. Then, we obtain in the steady state
\begin{equation}\label{dark state condition 01}
\left(S_{-}^{2}-N\alpha^{2}\right)\ket{D}\bra{D}S_{+}^{2}-S_{+}^{2}\left(S_{-}^{2}-N\alpha^{2}\right)\ket{D}\bra{D}+{\rm H.c.}=0,
\end{equation}
where $\ket{D}$ is the dark state of the ensemble, and
\begin{equation}
\alpha=i\sqrt{2\chi_{\rm 2at}/\kappa_{\rm 2at}}=i\sqrt{\Omega/\chi}.
\end{equation}
This indicates that the dark state $\ket{D}$ satisfies
\begin{equation}\label{seq:cat state condition}
\left(S_{-}^{2}-N\alpha^{2}\right)\ket{D}=0.
\end{equation}
We now express $\ket{D}$, in terms of the eigenstates $\ket{S=N/2, m_{z}=-N/2+n}$ of the collective spin operator $S_{z}$, as
\begin{equation}
\ket{D}=\sum_{n}c_{n}\ket{n},
\end{equation}
where, for simplicity, we have defined $\ket{n}\equiv\ket{S=N/2, m_z=-N/2+n}$. Here, $n$ refers to the number of excited atoms in the ensemble. The condition in Eq.~(\ref{seq:cat state condition}) gives two recursion relations as follows
\begin{align}\label{seq:relation for even cat state}
c_{2n+k}=\;\frac{\varepsilon^{n}}{\sqrt{(2n+k)!}}c_{k},
\end{align}
where $k=0,1$. Here, we have worked within the low-excitation regime, in which $\average{S_{z}}\approx -N/2$, such that the main contributions to the dark state $\ket{D}$ are from these components with $n\ll N$.

The recursion relation in Eq.~(\ref{seq:relation for even cat state}) reveals that, when the ensemble is initially in a collective spin state $\ket{n}$ with an even $n$, e.g., in the ground state $\ket{0}$ (i.e., a spin coherent state with all atoms in the ground state), the dark state $\ket{D}$ can be expressed as,
\begin{align}\label{even cat state_01}
\ket{D}_{\rm even}=\frac{1}{\sqrt{\cosh\left|\alpha\right|^{2}}}\sum_{n}\frac{\alpha^{2n}}{\sqrt{\left(2n\right)!}}\ket{2n}.
\end{align}
Similarly, when the ensemble is initially in a collective spin state $\ket{n}$ with an odd $n$, e.g., in the single-excitation state $\ket{1}$ (i.e.,  a state with only one atom is excited), the dark state $\ket{D}$ becomes
\begin{align}\label{odd cat state_01}
\ket{D}_{\rm odd}=\frac{1}{\sqrt{\sinh\left|\alpha\right|^{2}}}\sum_{n}\frac{\alpha^{2n+1}}{\sqrt{\left(2n+1\right)!}}\ket{2n+1}.
\end{align}

On the other hand, the spin coherent state $\ket{\theta, \phi}$ is defined as 
\begin{equation}\label{seq:coherent spin state_rotation operator}
\ket{\theta, \phi}=R\left(\theta,\phi\right)\ket{0}.
\end{equation}
Here, 
\begin{equation}\label{seq:rotation operation_factorization}
R\left(\theta,\phi\right)=\exp\left(\tau S_{+}\right)\exp\left[\ln\left(1+|\tau|^{2}\right)S_{z}\right]\exp\left(-\tau^{*}S_{-}\right),
\end{equation}
is a rotation operator with $\tau=\exp\left(i\phi\right)\tan\left(\theta/2\right)$.  
In the low-excitation limit, $S_{+}^{n}\ket{0}\approx\sqrt{n!N^{n}}\ket{n}$,
and then
\begin{equation}\label{seq:coheren spin state}
\ket{\theta,\phi}\approx\exp\left(-N\left|\tau\right|^{2}/2\right)\sum_{n}\frac{\left(\sqrt{N}\tau\right)^{n}}{\sqrt{n!}}\ket{n}.
\end{equation}
By setting $\sqrt{N}\tau=\alpha$, we further have
\begin{align}\label{seq:cat states}
\ket{D}_{\rm even,odd}=\mathcal{A}_{\pm}\left(\ket{\theta,\phi}\pm\ket{\theta,\phi+\pi}\right)=\ket{\mathcal{C}_{\pm}},
\end{align}
where $\mathcal{A}_{\pm}=1/\{2[1\pm \exp(-2\left|\alpha\right|^{2})]\}^{1/2}$. This is what we have already given in Eq.~(\ref{seq:atomic cat states in spin coherent states}).

We now consider the case when the atomic ensemble is initialized in a spin coherent  state $\ket{\theta_{0},\phi_{0}}$. In this case, the atomic ensemble evolves into a subspace spanned by the cat states $\left\{\ket{\mathcal{C}_{+}},\ket{\mathcal{C}_{-}}\right\}$ and, thus, its steady state is 
\begin{equation}
\rho_{\rm ens}^{\rm ss}=c_{++}\ket{\mathcal{C}_{+}}\bra{\mathcal{C}_{+}}+c_{--}\ket{\mathcal{C}_{-}}\bra{\mathcal{C}_{-}}+c_{+-}\ket{\mathcal{C}_{+}}\bra{\mathcal{C}_{-}}+c_{+-}^{*}\ket{\mathcal{C}_{-}}\bra{\mathcal{C}_{+}}.
\end{equation}
To obtain the amplitudes $c_{++}$, $c_{--}$, and $c_{+-}$, we follow the method in Refs.~\cite{S_albert2014symmetries,S_mirrahimi2014dynamically}, and after straightforward calculations, find that
\begin{align}
c_{++}=&\frac{1}{2}\left[1+\exp\left(-2\left|\alpha_{0}\right|^{2}\right)\right],\\
c_{--}=&1-c_{++}=\frac{1}{2}\left[1-\exp\left(-2\left|\alpha_{0}\right|^{2}\right)\right],\\
c_{+-}=&-\frac{\alpha_{0}^{*}\left|\alpha\right|\exp\left(-\left|\alpha_{0}\right|^{2}\right)}{\sqrt{2\sinh\left(2\left|\alpha\right|^{2}\right)}}\int_{0}^{\pi}d\varphi I_{0}\left(\left|\alpha^{2}-\alpha_{0}^{2}\exp\left(i2\varphi\right)\right|\right)\exp\left(-i\varphi\right),
\end{align}
where $\alpha_{0}=\sqrt{N}\exp\left(i\phi_{0}\right)\tan\left(\theta_{0}/2\right)$, and $I_{0}\left(\bigcdot\right)$ is the modified Bessel function of the first kind.

The above results show that the ensemble states are steered into a 2D quantum manifold spanned by the cat states $\ket{\mathcal{C}_{+}}$ and $\ket{\mathcal{C}_{-}}$. In typical atomic ensembles, spin relaxation is extremely weak, such that the dominant noise source is spin dephasing. However, the engineered two-atom decay can protect the cat states of the quantum manifold against spin dephasing. As a result, these cat states have a very long lifetime even with modest parameters, and thus, can be used for fundamental studies of quantum physics. Moreover, this atomic-cat-state manifold stabilized by the two-atom decay could also be used to encode logical qubits (i.e., cat qubits) for fault-tolerant quantum computation, as an alternative to the photonic-cat-state manifold stabilized by two-photon loss~\cite{S_mirrahimi2014dynamically}.

\subsection{S6. Strongly suppressed spin dephasing}
\label{Strongly suppressed spin dephasing}
\begin{figure}[t]
	\centering
	\includegraphics[width=13.0cm]{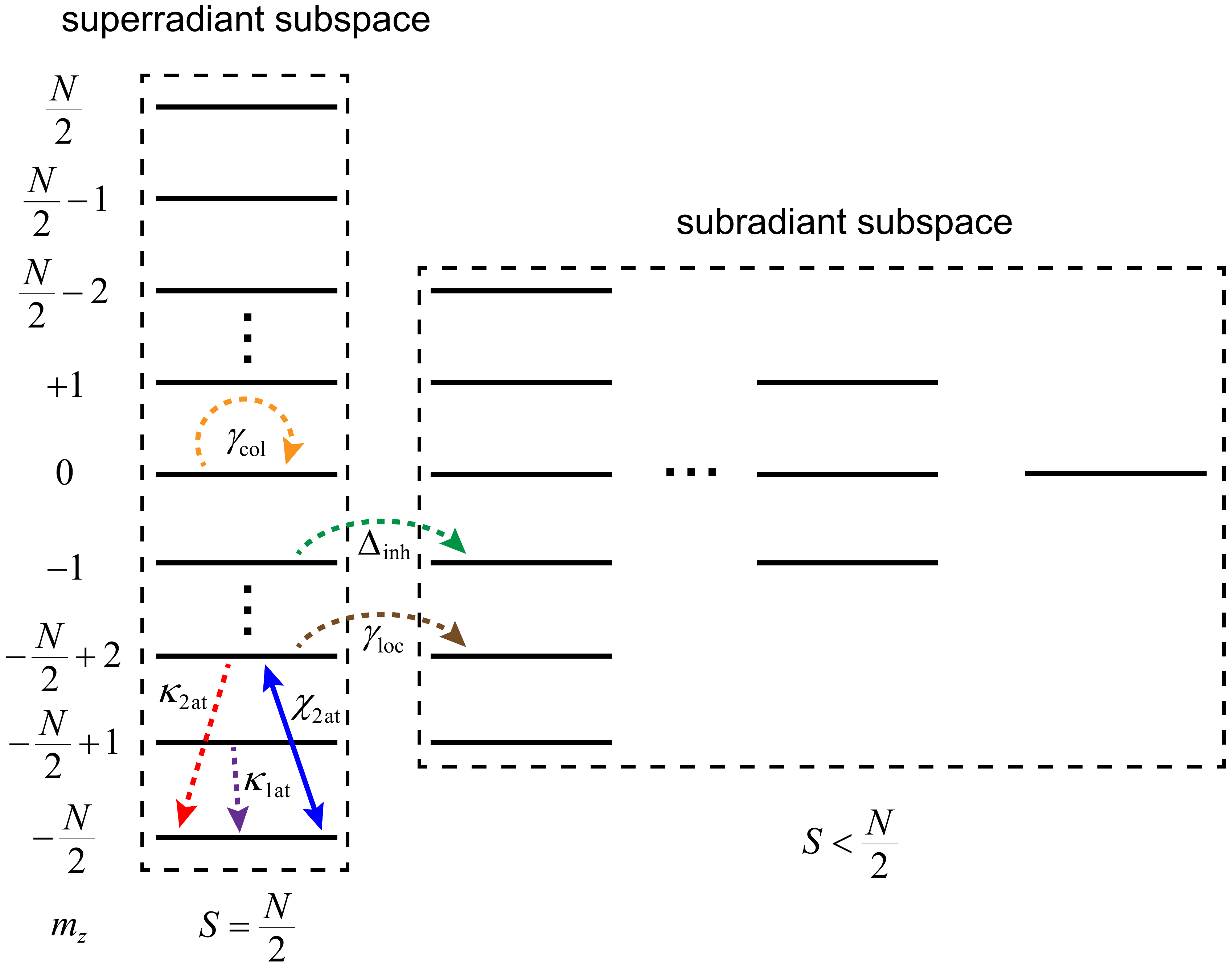}
	\caption{Dicke space for an ensemble consisting of $N$ identical two-level atoms or spins. The full space can be separated into the superradiant subspace of total spin $S=N/2$ and the subradiant subspace of total spin $S<N/2$. The blue solid double-headed arrow represents the two-atom excitation ($\chi_{\rm 2at}$), and the dashed arrows represent the dissipative processes, including single-atom decay ($\kappa_{\rm 1at}$), two-atom decay ($\kappa_{\rm 2at}$), collective dephasing ($\gamma_{\rm col}$), local dephasing ($\gamma_{\rm loc}$), and inhomogeneous broadening ($\Delta_{\rm inh}$). The two-atom decay and excitation only act inside the superradiant subspace, and thus the resulting cat states are stored inside this subspace. While collective dephasing does not couple the superradiant subspace to the subradiant subspace, local dephasing and inhomogeneous broadening couple these subspaces.}\label{fig_ensemble_space}
\end{figure}

In this section, we discuss the strong suppression of spin dephasing of atomic ensembles by the engineered two-atom decay. In general, the ensemble dephasing noise can be classified into three different types, i.e., collective spin dephasing, local spin dephasing, and inhomogeneous broadening. Below we show that as long as the rate $\gamma$ of convergence of cat states ($\gamma>|\alpha|^2\kappa_{\rm 2at}$) is much stronger than the collective dephasing rate $\gamma_{\rm col}$, the local dephasing rate $\gamma_{\rm loc}$, and the inhomogeneous linewidth $\Delta_{\rm inh}$, the engineered two-atom decay is capable of suppressing all of these dephasing processes. Here, the rate $\gamma$ describes how rapidly the steady cat states can be reached. As a result, steady cat states can be achieved with high fidelity.

To proceed, we note that our atomic cat states are stored in the superradiant subspace, rather than in the subradiant subspace. Here, the superradiant (subradiant) subspace refers to the manifold of total spin $S=N/2$ ($S<N/2$), as shown in Fig.~\ref{fig_ensemble_space}.



\subsection{A. Collective spin dephasing}

\begin{figure}[h]
	\centering
	\includegraphics[width=8.0cm]{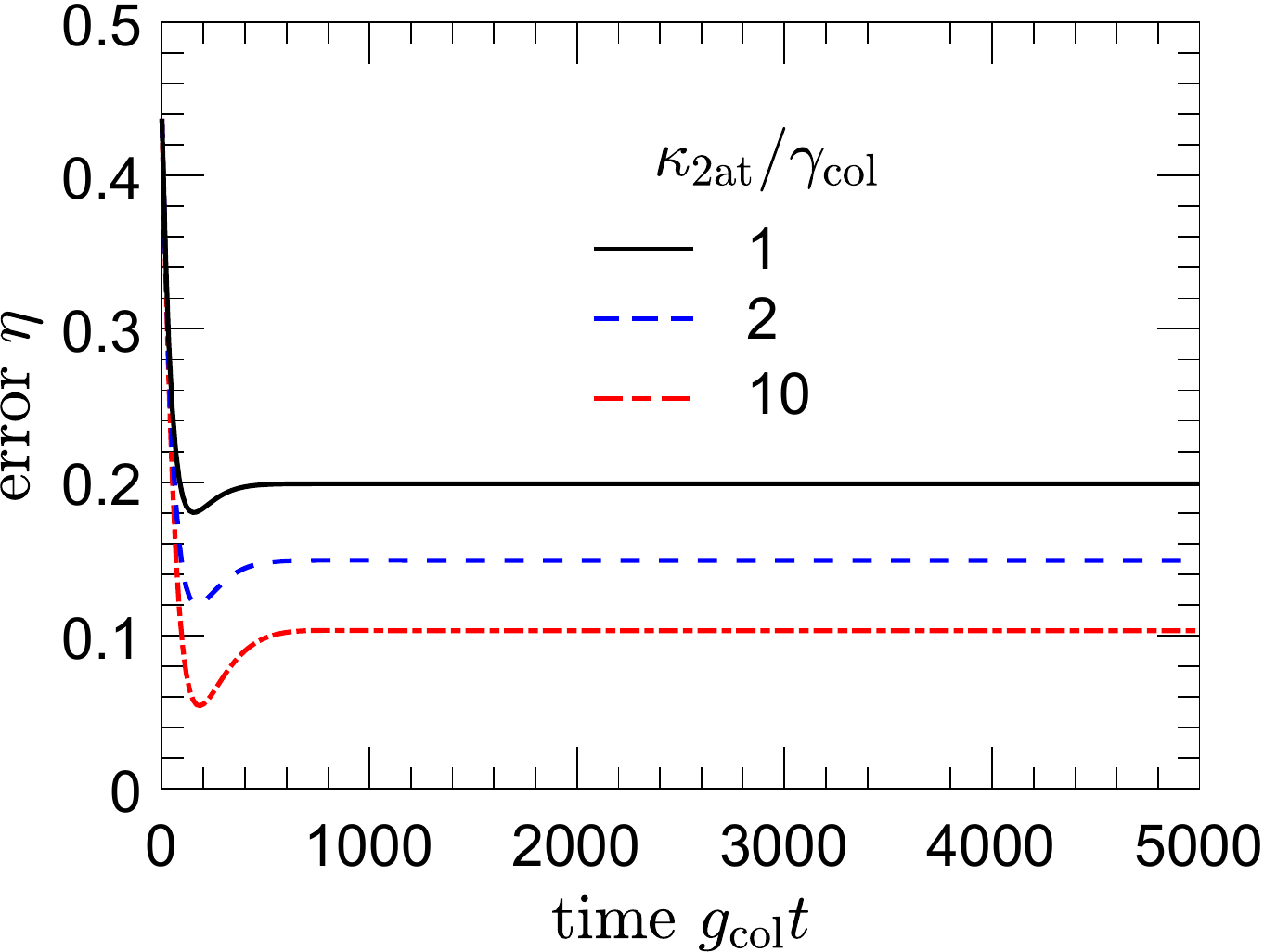}
	\caption{Effects of collective spin dephasing on the preparation error $\eta$ of the state $\ket{\mathcal{C}_{+}}$ of size $|\alpha|^{2}=2$. We integrated the effective master equation (4) in the main article, with an additional collective spin dephasing $\gamma_{\rm col}\sum_{j=1}^{N}\mathcal{L}(\sigma_{j}^{z})\rho_{\rm ens}$. For simplicity, we set $\kappa_{\rm 1at}=0$, so that only the effects of collective spin dephasing are shown. Other parameters are: $N=10$, $J=3g_{\rm col}$, $\delta_{p}=J^{2}/(20g_{\rm col})$, $\kappa_{p}=5\chi$, and $\kappa_{s}=0.3\kappa_{p}$.}\label{sfig_collective_dephasing}
\end{figure}

We first consider collective spin dephasing, which can be described with the Lindblad dissipator,
\begin{equation}\label{collective_dephasing}
\gamma_{\rm col}\mathcal{L}(S_{z})\rho_{\rm ens}=\gamma_{\rm col}\left(S_{z}\rho_{\rm ens}S_{z}-\frac{1}{2}S_{z}S_{z}\rho_{\rm ens}-\frac{1}{2}\rho_{\rm ens}S_{z}S_{z}\right).
\end{equation}
It arises when the atoms or spins of the ensemble are simultaneously coupled to a common bath. For example, the coupling to the collective phonon modes of the diamond can lead to collective dephasing for NV spin ensembles~\cite{S_prasanna2018cooperative}. Such a dephasing process does not couple the superradiant to subradiant subspace as shown in Fig.~\ref{fig_ensemble_space}, and as a result, the excitation-number parity of the superradiant subspace is {\it conserved}. Thus, collective spin dephasing can be suppressed by the two-atom decay, as long as the condition 
\begin{equation}
|\alpha|^2\kappa_{\rm 2at}\gg\gamma_{\rm col}
\end{equation}
is satisfied. This dissipative suppression can be better understood from the quantum-jump approach. The jump operator $S_{z}$, when acting, e.g., on the state $\ket{\mathcal{C}_{+}}$,  excites a state 
\begin{equation}
\ket{\psi}=\mathcal{A}_{+}\left[R\left(\theta,\phi\right)-R\left(\theta,\phi+\pi\right)\right]\ket{1},
\end{equation} 
according to 
\begin{equation}
S_{z}\ket{\mathcal{C}_{+}}=\left(-\frac{N}{2}+\left|\alpha\right|^2\right)\ket{\mathcal{C}_{+}}+\alpha\ket{\psi},
\end{equation}
where $\mathcal{A}_{+}=1/\{2[1+\exp(-2\left|\alpha\right|^{2})]\}^{1/2}$, and $R\left(\theta,\phi\right)$ is defined in Eq.~(\ref{seq:rotation operation_factorization}). It is seen that the state $\ket{\psi}$ still has even parity, and thus can be autonomously driven back to the target state $\ket{\mathcal{C}_{+}}$ by the two-atom decay. As shown in Fig.~\ref{sfig_collective_dephasing}, a steady cat state is achieved in the presence of collective spin dephasing.

\subsection{B. Local spin dephasing}

We now consider local spin dephasing, described by the Lindblad dissipator
\begin{equation}\label{local_dephasing}
\gamma_{\rm loc}\sum_{j=1}^{N}\mathcal{L}\left(\sigma_{j}^{z}\right)\rho_{\rm ens}=\gamma_{\rm loc}\sum_{j=1}^{N}(\sigma_{j}^{z}\rho_{\rm ens}\sigma_{j}^{z}-\rho_{\rm ens}).
\end{equation}
The quantum jump, $\sigma_{j}^{z}$, when acting on the superradiant state $\ket{n}\equiv\ket{S=N/2, m_{z}=-N/2+n}$, results in a superposition of the state $\ket{n}$ with a subradiant state $\ket{n}^{\perp}_{j}$. This indicates a dissipative coupling of the superradiant to subradiant subspace, as shown in Fig.~\ref{fig_ensemble_space}, yielding
\begin{equation}\label{eq:sigma_z_on_n}
\sigma_{j}^{z}\ket{n}=\left(1-\frac{2n}{N}\right)\ket{n}-2\sqrt{\frac{n}{N}}\ket{n}^{\perp}_{j}.
\end{equation}
Here, the subradiant state $\ket{n}^{\perp}_{j}$ is orthogonal to the superradiant state $\ket{n}$ and has the same magnetic quantum number $m_{z}$ as $\ket{n}$.

\begin{figure}[t]
	\centering
	\includegraphics[width=8.0cm]{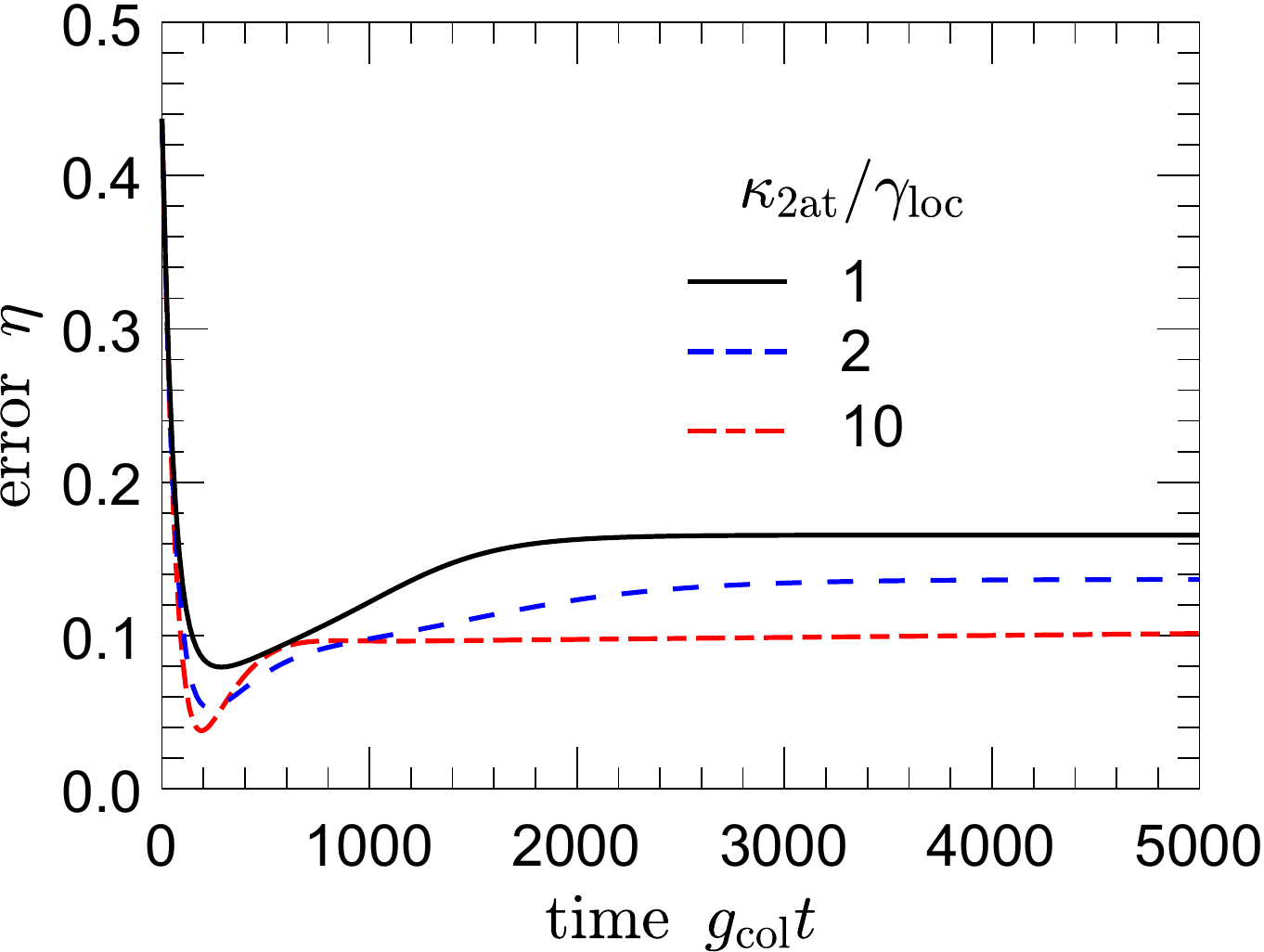}
	\caption{Effects of local spin dephasing on the preparation error $\eta$ of the state $\ket{\mathcal{C}_{+}}$ of size $|\alpha|^{2}=2$. We integrated the effective master equation (4) in the main article, with an additional local spin dephasing $\gamma_{\rm loc}\sum_{j=1}^{N}\mathcal{L}(\sigma_{j}^{z})\rho_{\rm ens}$. For simplicity, we set $\kappa_{\rm 1at}=0$, so that only the effects of local spin dephasing are shown. Other parameters are set the same as in Fig.~\ref{sfig_collective_dephasing}.}\label{fig_local_dephasing}
\end{figure}

As an example, we consider the action of the quantum jump $\sigma_{j}^{z}$ on the even cat state $\ket{\mathcal{C}_{+}}$. Note that similar results hold for the odd cat state $\ket{\mathcal{C}_{-}}$. According to Eq.~(\ref{eq:sigma_z_on_n}), we obtain
\begin{equation}\label{jump_sigma_z_j}
\sigma_{j}^{z}\ket{\mathcal{C}_{+}}=\sum_{n}c_{2n}\sigma_{j}^{z}\ket{2n}=\sum_{n}c_{2n}\left(1-\frac{4n}{N}\right)\ket{2n}-2\sum_{n}c_{2n}\sqrt{\frac{2n}{N}}\ket{2n}^{\perp}_{j}.
\end{equation}
It is seen that the quantum jump $\sigma_{j}^{z}$ distorts the cat state $\ket{\mathcal{C}_{+}}$, but {\it conserves} the excitation-number parity of the superradiant subspace, although it carries some information about the cat state $\ket{\mathcal{C}_{+}}$ away from the superradiant to subradiant subspace. Thus as long as 
\begin{equation}
|\alpha|^{2}\kappa_{\rm 2at}\gg\gamma_{\rm loc},
\end{equation}
the two-atom decay and excitation, which act only inside the superradiant subspace (see Fig.~\ref{fig_ensemble_space}), can autonomously steer the dephasing-distorted cat state [i.e., the superradiant component $\sum_{n}c_{2n}\left(1-4n/N\right)\ket{2n}$] back to the target state $\ket{\mathcal{C}_{+}}$.
This indicates that, as confirmed in Fig.~\ref{fig_local_dephasing}, local spin dephasing can be strongly suppressed and, consequently, that a steady cat state can be achieved in the superradiant subspace.

\subsection{C. Inhomogeneous broadening}

Let us now consider inhomogeneous broadening of the ensemble. Its detrimental effects can, in principle, be completely canceled by spin-echo pulses~\cite{S_hahn1950spin}. For ensembles of ultracold atoms~\cite{S_hattermann2017coupling}, these detrimental effects can also be minimized through spin self-rephasing collisions, even without the need for spin-echo pulse sequences~\cite{S_deutsch2010spin,S_kleine2011extended}.

\begin{figure}[b]
	\centering
	\includegraphics[width=8.0cm]{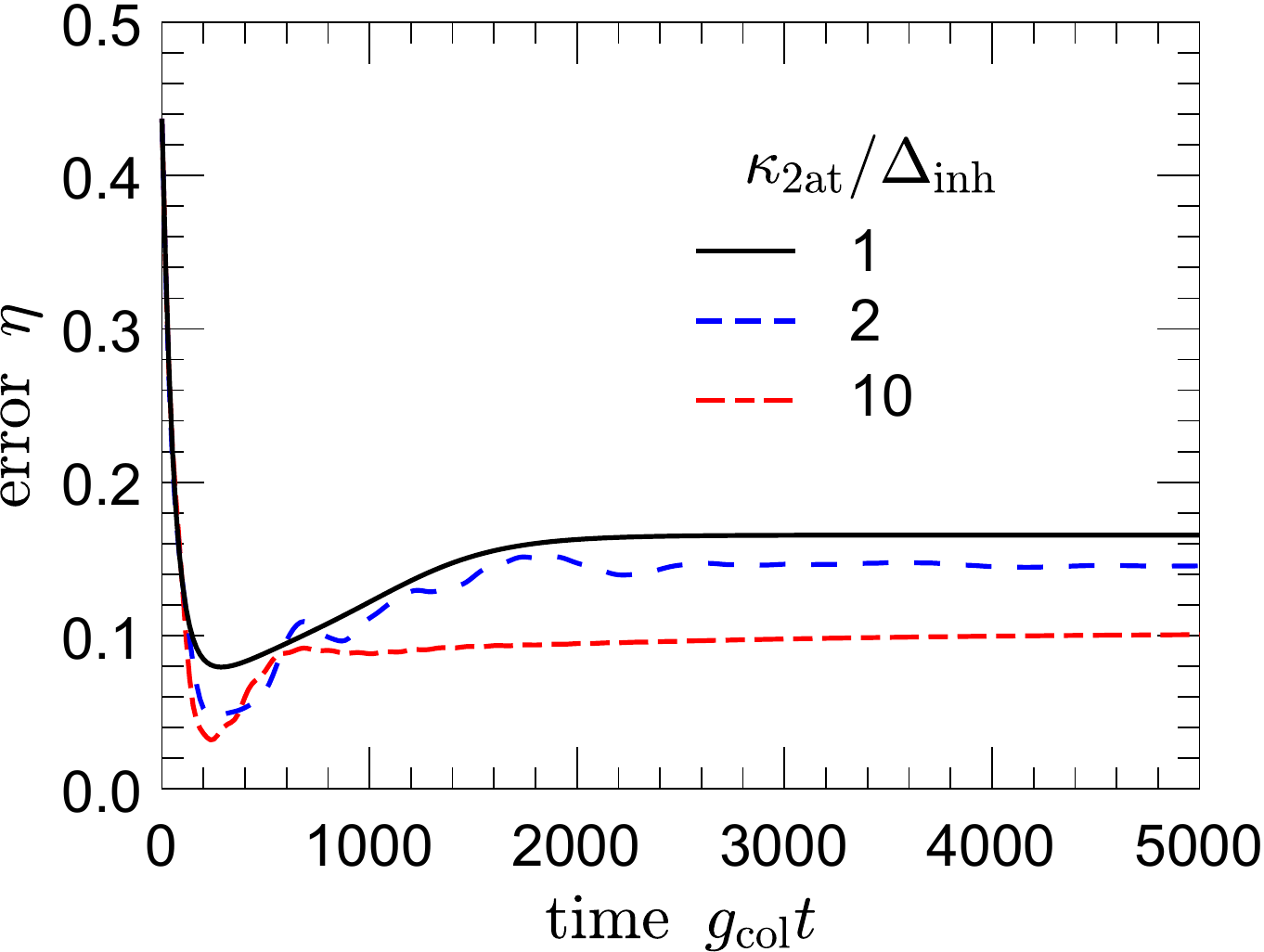}
	\caption{Effects of inhomogeneous broadening on the preparation error $\eta$ of the state $\ket{\mathcal{C}_{+}}$ of size $|\alpha|^{2}=2$. We integrated the effective master equation (4) in the main article, with an additional inhomogeneous broadening $H_{\rm inh}=\frac{1}{2}\sum_{j=1}^{N}\delta_{j}\sigma_{j}^{z}$. For simplicity, we set $\kappa_{\rm 1at}=0$, so that only the effects of inhomogeneous broadening are shown. The frequency shifts $\delta_{j}$ are randomly given according to a Lorentzian distribution of linewidth $\Delta_{\rm inh}$. Other parameters are set the same as in Fig.~\ref{sfig_collective_dephasing}.}\label{fig_inh}
\end{figure}

The Hamiltonian modeling inhomogeneous broadening is given by
\begin{equation}\label{eq:inhomogeneous_Hamiltonian}
H_{\rm inh}=\frac{1}{2}\sum_{j=1}^{N}\delta_{j}\sigma_{j}^{z},
\end{equation}
where $\delta_{j}=\omega_{j}-\omega_{q}$. Here, $\omega_{j}$ is the transition frequency of the $j$th qubit spin, and $\omega_{q}$ can be viewed as the average of transition frequencies of all the qubit spins. Under the time evolution, each constituent of the symmetric superradiant state $\ket{n}$ acquires a random phase originating from inhomogeneous broadening. As a result, the superradiant state $\ket{n}$ is coupled to a subradiant state as shown in Fig.~\ref{fig_ensemble_space}, thus destroying the cat states.

Nevertheless, according to the action of the operator $\sigma_{j}^{z}$ on the cat state $\ket{\mathcal{C}_{+}}$, as given in Eq.~(\ref{jump_sigma_z_j}), inhomogeneous broadening {\it conserves} the excitation-number parity of the superradiant subspace. Thus, the two-atom decay can strongly suppress inhomogeneous broadening when 
\begin{equation}
|\alpha|^{2}\kappa_{\rm 2at}\gg\Delta_{\rm inh}.
\end{equation}

To confirm the suppression of inhomogeneous broadening, we perform numerical simulations, as shown in Fig.~\ref{fig_inh}. Inhomogeneous broadening is assumed to be, as an example, the Lorentzian distribution with a width $\Delta_{\rm inh}$, and similar results hold for other spectra. It is seen from Fig.~\ref{fig_inh} that a cat state is stabilized in the presence of inhomogeneous broadening, as expected.


\subsection{D. Total effects of collective dephasing, local dephasing, and inhomogeneous broadening}

\begin{figure}[h]
	\centering
	\includegraphics[width=8.0cm]{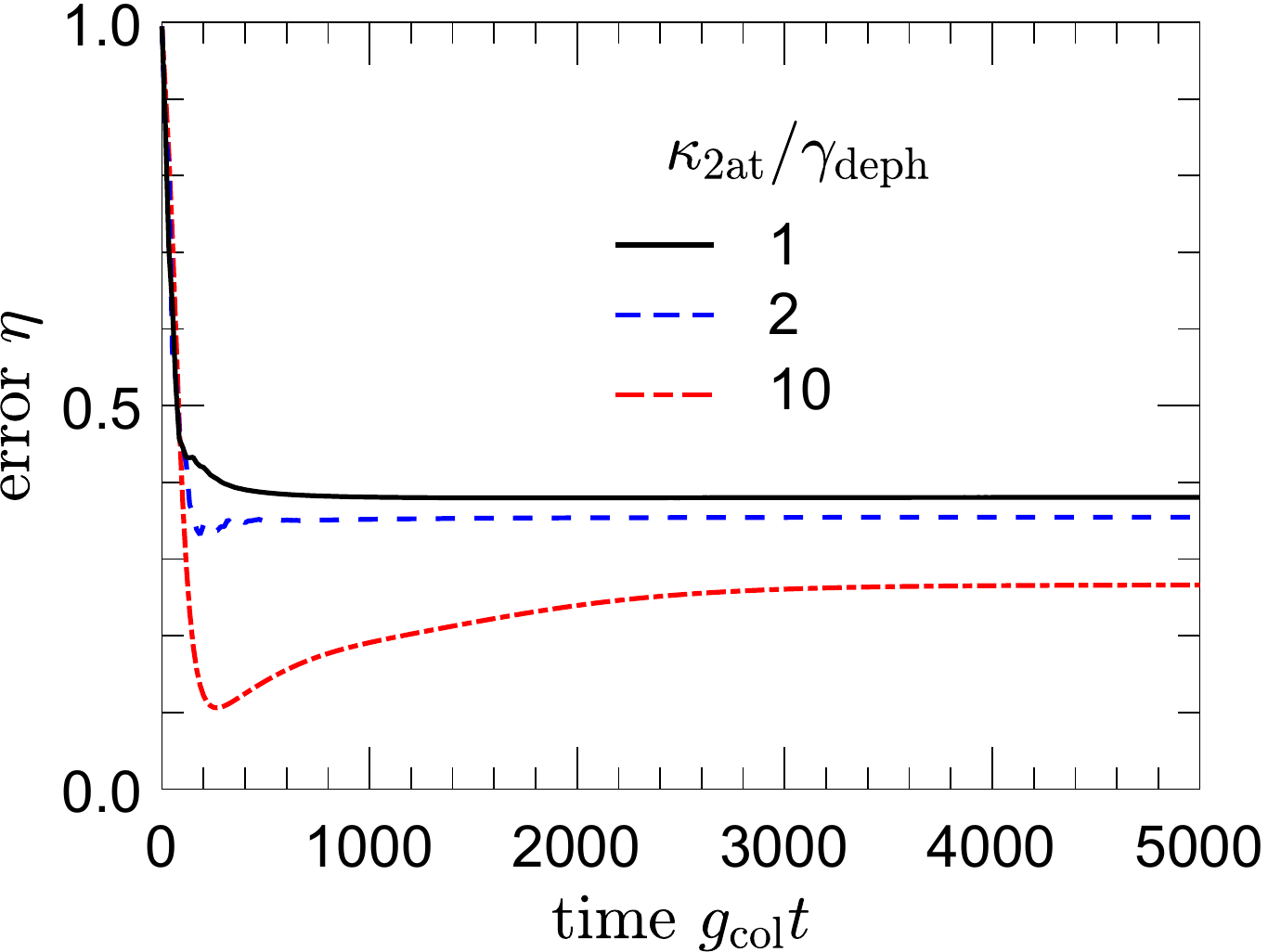}
	\caption{Total effects of collective dephasing, local dephasing, and inhomogeneous broadening on the preparation error $\eta$ of the state $\rho_{\rm ens}^{\rm ss}$ of size $|\alpha|^{2}=2$. We integrated the effective master equation (4) in the main article, with an additional spin dephasing $\gamma_{\rm col}\mathcal{L}(S_{z})\rho_{\rm ens}$, local spin dephasing $\gamma_{\rm loc}\sum_{j=1}^{N}\mathcal{L}(\sigma_{j}^{z})\rho_{\rm ens}$, and inhomogeneous broadening $\frac{1}{2}\sum_{j=1}^{N}\delta_{j}\sigma_{j}^{z}$. The frequency shifts $\delta_{j}$ are randomly given according to a Lorentzian distribution of linewidth $\Delta_{\rm inh}$. For simplicity, we set $\gamma_{\rm col}=\gamma_{\rm loc}=\Delta_{\rm inh}\equiv\gamma_{\rm deph}$, and $\kappa_{\rm 1at}=0$, so that only the effects of these dephasing processes are shown. Other parameters are set the same as in Fig.~\ref{sfig_collective_dephasing}.}\label{fig_inh_local_col_dephasing}
\end{figure}

In Fig.~\ref{fig_inh_local_col_dephasing}, we show the total effects of collective dephasing, local dephasing, and inhomogeneous broadening on the superposition, $\rho_{\rm ens}^{\rm ss}$, of the even and odd cat states, as a supplement to Fig. 3(a) in the main article which shows the case of the state $\ket{\mathcal{C}_{+}}$. As expected, the steady 2D cat-state manifold can be obtained, even when these three sources of dephasing noise are present simultaneously.

Note that in Fig.~\ref{fig_inh_local_col_dephasing}, the preparation error $\eta$, especially for the $\kappa_{\rm 2at}=10\gamma_{\rm deph}$ case, is limited by the small number $N$ which is chosen for the convenience of numerical simulations. Here, we have assumed that $\gamma_{\rm deph}\equiv\gamma_{\rm col}=\gamma_{\rm loc}=\Delta_{\rm inh}$. A larger $N$ leads to a smaller $\eta$, until the bosonic approximation is valid well, i.e., until the collective behavior of the ensemble can be well approximated by a harmonic oscillator. A similar increase in $\eta$ can also be observed in Figs.~\ref{sfig_collective_dephasing}, ~\ref{fig_local_dephasing}, and~\ref{fig_inh}.

\subsection{S7. Inhomogeneous broadening in nitrogen-vacancy center ensembles}
\label{Inhomogeneous broadening in nitrogen-vacancy center ensembles}

In Sec.~\hyperref[Strongly suppressed spin dephasing]{S6}, we have discussed three types of dephasing noise for our model. Different types of atomic or spin ensembles have different dephasing mechanisms. Below, we take ensembles of nitrogen-vacancy (NV) center electron spins in diamond, as an example, to discuss the source of dephasing. In these systems, the source of dephasing is inhomogeneous broadening of the NV transition.

The electronic
ground state of NV centers is a spin triplet, which has $m_s = 0$
and $\pm1$ sublevels. We use $\ket{0}$ and $\ket{\pm1}$ to label these three sublevels. The zero-field splitting between the states $\ket{0}$ and
$\ket{\pm1}$ is $\sim2.87$~GHz. In the presence of a static field,
the degenerate states $\ket{\pm1}$ are split with a Zeeman splitting $\Delta_{\rm zm}$.
In order to encode a two-level atom or qubit here, we assume that the state $\ket{0}$ is used as the ground state
and the state $\ket{+1}$ as the excited state. Inhomogeneous broadening of the spin transition can be described by the Hamiltonian in Eq.~(\ref{eq:inhomogeneous_Hamiltonian}), which for convenience is recalled here
\begin{equation}
H_{\rm inh}=\frac{1}{2}\sum_{j=1}^{N}\delta_{j}\sigma_{j}^{z},
\end{equation}
where $\sigma_{j}^{z}=\ket{+1}_{j}\bra{+1}-\ket{0}_{j}\bra{0}$, and $\delta_{j}=\omega_{j}-\omega_{q}$. Here, $\omega_{j}$ is the transition frequency of the $j$th qubit spin, and $\omega_{q}$ can be viewed as the average of transition frequencies of all the qubit spins.

In general, inhomogeneous broadening of NV ensembles originates from the interactions of the NV centers with (A) the local strain field, (B) the $^{13}$C and $^{14}$N nuclear spins, and (C) the P1 centers. Thus, the frequency shift $\delta_{j}$ can be separated into three parts, i.e.,
\begin{equation}
\delta_{j}=\delta_{j}^{\rm str}+\delta_{j}^{\rm nuc}+\delta_{j}^{\rm P1},
\end{equation}
which includes contributions from the strain field ($\delta_{j}^{\rm str}$), the $^{13}$C and $^{14}$N nuclear spins ($\delta_{j}^{\rm nuc}$), and the P1 centers ($\delta_{j}^{\rm P1}$).



\subsection{A. Local strain}

The local strain field breaks the $C_{3v}$ symmetry of the NV center, and as a result shifts the frequency of the states $\ket{\pm1}$. The NV electronic spin is coupled to the local strain field via the Hamiltonian~\cite{S_doherty2013nitrogen}
\begin{align}\label{eq:strain_Hamiltonian}
H_{\rm strain}=\;&d_{\parallel}\mathcal{E}_{\rm str}^{z}\mathcal{S}_{z}^{2}+d_{\perp}\mathcal{E}_{\rm str}^{x}\left(\mathcal{S}_{y}^{2}-\mathcal{S}_{x}^{2}\right)+d_{\perp}\mathcal{E}_{\rm str}^{y}\left(\mathcal{S}_{x}\mathcal{S}_{y}+\mathcal{S}_{y}\mathcal{S}_{x}\right)\nonumber\\
=\;&\Pi_{z}(\ket{+1}\bra{+1}+\ket{-1}\bra{-1})+\left(\Pi_{\perp}\ket{+1}\bra{-1}+{\rm H.c.}\right),
\end{align}
where $\Pi_{z}=d_{\parallel}\mathcal{E}_{\rm str}^{z}$, $\Pi_{\perp}=-d_{\perp}\left(\mathcal{E}_{\rm str}^{x}+i\mathcal{E}_{\rm str}^{y}\right)$, and $\vec{\mathcal{S}}=(\mathcal{S}_{x},\mathcal{S}_{y},\mathcal{S}_{z})$ is the NV spin operator. Here, $\vec{\mathcal{E}}_{\rm str}=(\mathcal{E}_{\rm str}^{x},\mathcal{E}_{\rm str}^{y},\mathcal{E}_{\rm str}^{z})$ represents the strain field, and $d_{\parallel}\sim2\pi\times0.35$~Hz cm/V, $d_{\perp}\sim2\pi\times17$~Hz cm/V are the axial and non-axial components of the ground-state electric dipole moment. The first term in Eq.~(\ref{eq:strain_Hamiltonian}) corresponds to the frequency shifts of the states $\ket{\pm1}$, and the second term describes their coupling. Due to the Zeeman splitting $\Delta_{\rm zm}$, the coupling between the states $\ket{\pm1}$ becomes largely detuned. As a result, the transition frequency of the qubit spin (i.e., the transition $\ket{0}\rightarrow\ket{+1}$) is shifted by
\begin{equation}
\delta^{\rm str}=\Pi_{z}+\frac{\left|\Pi_{\perp}\right|^{2}}{\Delta_{\rm zm}}.
\end{equation}
For a realistic parameter $|\Pi_{\perp}|=2\pi\times5$~MHz~\cite{S_kubo2010strong}, and a common Zeeman splitting $\Delta_{\rm zm}=2\pi\times100$~MHz, an estimate of $\delta^{\rm str}$ is therefore given by $\delta^{\rm str}\sim2\pi\times0.3$~MHz.

\subsection{B. Nuclear spins}

Natural diamond samples consist of $\sim98.9\%$ spinless $^{12}$C atoms and $\sim1.1\%$ $^{13}$C isotopes of nuclear spin $\mathcal{I}_{\rm C}=1/2$. These $^{13}$C atoms are randomly distributed in the diamond lattice. Moreover, the $^{14}$N  atoms constituting the NV centers each have a nuclear spin $\mathcal{I}_{\rm N}=1$. The NV centers are coupled to these $^{14}$N and $^{13}$C nuclear spins through hyperfine interactions, given by
\begin{equation}
H_{\rm nuc}=\vec{\mathcal{S}}\cdot\mathbb{A}_{\rm N}\cdot\vec{\mathcal{I}}_{N}+\vec{\mathcal{S}}\cdot\sum_{j}\mathbb{A}_{\rm C}\cdot\vec{\mathcal{I}}_{\rm C}^{j},
\end{equation}
where $\vec{\mathcal{I}}_{N}$ and $\vec{\mathcal{I}}_{\rm C}^{j}$ are the spin operators for the $^{14}$N atom and the $j$th $^{13}$C atom, respectively, while  $\mathbb{A}_{\rm N}$ and $\mathbb{A}_{\rm C}^{j}$ are the corresponding hyperfine interaction tensors. Working under the secular approximation, i.e., neglecting the $\mathcal{S}_{x}$ and $\mathcal{S}_{y}$ terms, the Hamiltonian $H_{\rm nuc}$ becomes approximated by
$H_{\rm nuc}\approx\delta^{\rm nuc}\mathcal{S}_{z}$~\cite{S_dreau2012high,S_zou2014implementation}, with
\begin{equation}\label{eq:interactions_with_nuclear_spins}
\delta^{\rm nuc}=\mathbb{A}_{\rm N}m_{N}+\sum_{j}\mathbb{A}_{\rm C}^{j}m_{\rm C}^{j},
\end{equation}
where $m_{\rm N}=0$, $\pm1$ and $m_{\rm C}^{j}=\pm1/2$ are magnetic quantum numbers.

The coupling to the $^{14}$N nuclear spin splits the state $\ket{+1}$ (or $\ket{-1}$) into three hyperfine sublevels, equally spaced by $\mathbb{A}_{\rm N}\sim2\pi\times2.16$~MHz. This results in a linewidth broadening $\sim2\pi\times4.3$~MHz.

The $^{13}$C hyperfine splitting depends on the positions of the $^{13}$C nuclear spins relative to the NV center. According to the studies in Ref.~\cite{S_zhao2012decoherence}, the coherence time induced by the $^{13}$C hyperfine coupling is $\sim 2$~$\mu$s, implying a linewidth broadening of $\sim2\pi\times 80$~kHz. If $^{12}$C-enriched methane is used as a carbon source to prepare the diamond samples~\cite{S_grezes2015storage}, then the concentration of $^{13}$C nuclear spins (and as a result the corresponding linewidth broadening) can be significantly reduced.

\subsection{C. P1 centers}

In diamond samples, single substitutional nitrogen atoms (so-called P1 centers), which were not converted into the NV centers, are the main paramagnetic impurities and each of them has an unpaired electron. These inevitable impurities form an electron spin bath, and the NV center is coupled to it through the dipole-dipole interaction, which is described by the Hamiltonian~\cite{S_taylor2008high}:
\begin{align}
H_{\rm P1}=\sum_{j}\frac{\mu_{0}g_{s}^{2}\mu_{B}^{2}}{4\pi \left|\vec{r}_{j}\right|^{3}}\mathcal{S}_z\left[\vec{n}_{z}-3\left(\vec{n}_{z}\cdot\vec{n}_{j}\right)\vec{n}_{j}\right]\cdot \vec{\mathcal{S}}_{j},
\end{align}
where $\vec{\mathcal{S}}_{j}$ is the bath spin located at position $\vec{r}_{j}$, and $\vec{n}_{j}=\vec{r}_{j}/|\vec{r}_{j}|$. In most experiments implementing the strong coupling of a high-density NV ensemble to a superconducting resonator~\cite{S_kubo2010strong,S_amsuss2011cavity,S_kubo2011hybrid,S_PhysRevA.85.012333,S_grezes2014multimode,S_putz2014protecting,S_astner2017coherent}, the residual P1 centers are the main source of decoherence of the NV ensemble, and a typical linewidth broadening is $\delta^{\rm P1}\sim2\pi\times7$~MHz~\cite{S_kubo2010strong,S_kubo2011hybrid,S_putz2014protecting}.

A solution to reduce the inhomogeneous linewidth induced by the P1 centers is to improve the efficient conversion of the P1 centers to the NV centers. The inhomogeneous linewidth would therefore be dominated by the hyperfine interaction with the $^{14}$N nuclear spin [i.e., the first term on the right-hand side of  Eq.~(\ref{eq:interactions_with_nuclear_spins})]. That is, the inhomogeneous linewidth would be limited to $\sim2\pi\times4.3$~MHz, as experimentally reported in Refs.~\cite{S_grezes2015storage,S_angerer2018superradiant}.

\subsection{D. Short summary}
The detrimental effects of inhomogeneous broadening mentioned above are reversible and can in principle be completely eliminated by spin-echo techniques or dynamical decoupling pulse sequences. The residual inhomogeneous broadening can be further suppressed by the engineered two-atom decay in our proposal (see Sec.~\hyperref[Strongly suppressed spin dephasing]{S6}). Note that although we discuss the ensembles of NV spins, our model is generic and can be implemented with other types of ensembles, e.g., ensembles of trapped ultracold atoms. Inhomogeneous broadening of ultracold-atom ensembles, which arises mainly due to the trapping potential and the atomic interactions~\cite{S_hattermann2017coupling}, can be strongly reduced through spin self-rephasing collisions without the use of spin-echo or dynamical decoupling pulses~\cite{S_deutsch2010spin,S_kleine2011extended}.

\subsection{S8. Spin relaxation, thermal noise, and the maximum cat state lifetime}
\label{sec:Spin relaxation, thermal noise, and the maximum cat state lifetime}
In the main article, we discussed the effects of spin dephasing, and also showed that it can be strongly suppressed by the engineered two-atom decay. In this section, let us consider the effects of spin relaxation and thermal noise, and also the maximum cat state lifetime limited by them. Here, we proceed with the bosonic approximation. Such an approximation maps the spin coherent states $\ket{\theta, \phi}$ and $\ket{\theta, \phi+\pi}$ to the bosonic coherent states $\ket{\pm\alpha}$, respectively. Correspondingly, the cat states $\ket{\mathcal{C}_{\pm}}=\mathcal{A}_{\pm}\left(\ket{\theta,\phi}\pm\ket{\theta,\phi+\pi}\right)$ become $
\ket{\mathcal{C}_{\pm}}=\mathcal{A}_{\pm}\left(\ket{\alpha}\pm\ket{-\alpha}\right)$, as given in Eq.~(\ref{seq:atomic cat states in bosonic coherent states}).

Spin relaxation and thermal noise can be described by the Lindblad dissipators, $\gamma_{\rm relax}(n_{\rm th}+1)\mathcal{L}\left(b\right)\rho$ and $\gamma_{\rm relax}n_{\rm th}\mathcal{L}\left(b^{\dagger}\right)\rho$. Here, $\gamma_{\rm relax}$ is the spin relaxation rate, and $n_{\rm th}=\left[\exp\left(\hbar\omega_{q}/k_{B}T\right)-1\right]^{-1}$ is the thermal average boson number at temperature $T$. 
The Purcell decay rate of the cat state coherence, which is induced by single-photon loss of the signal cavity, is given by
\begin{equation}
\Gamma_{\rm 1at}=2\left|\alpha\right|^{2}\kappa_{\rm 1at},
\end{equation}
with $\kappa_{\rm 1at}=\left(g_{\rm col}/\Delta\right)^{2}\kappa_{s}$ as given in Eq.~(\ref{seq:induced single atom decay}).
At the same time, for a thermal background at $T\neq0$, an additional decay rate of the cat state coherence, which is induced by spin relaxation and thermal noise, is given by~\cite{S_kim1992schrodinger}
\begin{equation}
\Gamma_{\rm relax}=\left[2\left|\alpha\right|^{2}\left(1+2n_{\rm th}\right)+2n_{\rm th}\right]\gamma_{\rm relax}.
\end{equation}
By assuming realistic parameters $\omega_{q}=2\pi\times3$~GHz, $T=100$~mK, $\left|\alpha\right|^2=4$, and $\gamma_{\rm relax}=2\pi\times4$~mHz~\cite{S_amsuss2011cavity,S_grezes2014multimode}, we have
$\Gamma_{\rm relax}\approx2\pi\times54$~mHz, much smaller the decay rate, $\Gamma_{\rm 1at}\approx2\pi\times8.0$~Hz, which is obtained with $\kappa_{s}=2\pi\times10$~kHz and $\Delta/g_{\rm col}=100$. 
This means that the effects of both spin relaxation and thermal noise on the cat states $\ket{\mathcal{C}_{\pm}}$ can be safely neglected. In this case, the lifetime of these cat states is determined only by the Purcell single-atom decay rate $\kappa_{\rm 1at}$, and is given by
\begin{equation}\label{seq:atomic cat lifetime}
\tau_{\rm at}=\Gamma_{\rm 1at}^{-1}=
\left(\frac{\Delta}{g_{\rm col}}\right)^{2}\frac{1}{2\left|\alpha\right|^{2}\kappa_{s}}.
\end{equation} 
On the other hand, the intracavity photonic cat states $\ket{\mathcal{C}_{\pm}}_{\rm ph}$ in Eq.~(\ref{seq:photonic cat state}) mainly suffer from single-photon loss, e.g, with a rate $\kappa_{s}$, and thus their lifetime is given by~\cite{S_haroche2006exploring}, 
\begin{equation}\label{seq:photonic cat lifetime}
\tau_{\rm ph}=\frac{1}{2\left|\alpha\right|^{2}\kappa_{s}}.
\end{equation}
It is found from Eqs.~(\ref{seq:atomic cat lifetime}) and~(\ref{seq:photonic cat lifetime}) that $\tau_{\rm at}$ is longer than $\tau_{\rm ph}$ by a factor of $\left(\Delta/g_{\rm col}\right)^{2}$, i.e.,
\begin{equation}\label{seq:atomic cat lifetime_2}
\frac{\tau_{\rm at}}{\tau_{\rm ph}}=
\left(\frac{\Delta}{g_{\rm col}}\right)^{2}.
\end{equation}
According to the analysis in the main article, the factor $\left(\Delta/g_{\rm col}\right)^{2}$ can be tuned to be $\sim10^{4}$ under modest parameters. This indicates that \emph{the lifetime of our atomic cat states is longer than that of intracavity photonic cat states by up to four orders of magnitude for cat sizes of $\left|\alpha\right|^2\geq4$.} 

In fact, the decoherence rate $\Gamma_{\rm 1at}$ can be further decreased with the smaller single-photon loss rate $\kappa_{s}$ (i.e., the longer $T_{c}$). This results in a longer cat state lifetime. When $\Gamma_{\rm 1at}$ is comparable to or even smaller than $\Gamma_{\rm relax}$, the lifetime $\tau_{\rm at}$ is given by
\begin{equation}\label{seq:lifetime02}
\tau_{\rm at}=\left(\Gamma_{\rm 1at}+\Gamma_{\rm relax}\right)^{-1}.
\end{equation}
For a single-photon loss rate of $\kappa_{s}/2\pi=30$~Hz, we have $\Gamma_{\rm 1at}=2\pi\times24$~mH, which is smaller than $\Gamma_{\rm relax}\sim2\pi\times54$~mHz. In this case, Eq.~(\ref{seq:lifetime02}) gives a cat state lifetime of $\tau_{\rm at}\sim2$~sec. Ultimately, when decreasing the rate $\kappa_{s}$, the lifetime $\tau_{\rm at}$ increases to its maximum value,
\begin{equation}
\tau_{\rm at}^{\rm max}=\Gamma_{\rm relax}^{-1}.
\end{equation}
Using the parameters given above, we can predict a maximum lifetime of $\tau_{\rm at}^{\rm max}\sim3$~sec.


%

\end{document}